\documentclass{nejsds_ST}

\usepackage{amsmath} 
\usepackage{amssymb}
\usepackage{enumerate}
\usepackage{multirow}

\usepackage{algorithm2e} 

\usepackage[mathscr]{eucal}
\usepackage{times}

\usepackage{graphicx} 
\graphicspath{ {Images/} }
\usepackage{epstopdf}
\usepackage{subfig} 
\usepackage{wrapfig}

\usepackage{hypcap}

\usepackage{color}

\usepackage{subfiles} 

\renewcommand{\pmb}{{}}

\def\benu{\begin{enumerate}}
	\def\eenu{\end{enumerate}}
\newcommand{\bel}{\begin{eqnarray}\label}
    \newcommand{\eel}{\end{eqnarray}}
\newcommand{\bes}{\begin{eqnarray*}}
	\newcommand{\ees}{\end{eqnarray*}}
\newcommand{\bei}{\begin{itemize}}
	\newcommand{\eei}{\end{itemize}}

\def\veps{\varepsilon}

\def\htheta{\hat{\theta}_{\text{S}}}

\def\P{{\cal P}}									

\def\ftil{\widetilde{f}}

\def\bs{\pmb{s}}

\def\bx{\pmb{x}}

\def\bS{\pmb{S}}

\def\btheta{\pmb{\theta}}
\articletype{research-article}

\theoremstyle{remark}
\newtheorem{thm}{Theorem}

\newtheorem{lemma}{Lemma}
\newtheorem{cor}{Corollary}

\newtheorem{condition}{Condition}

\makeatletter
\newcommand{\distas}[1]{\mathbin{\overset{#1}{\kern\z@\sim}}}
\newsavebox{\mybox}\newsavebox{\mysim}
\newcommand{\distras}[1]{
	\savebox{\mybox}{\hbox{\kern3pt$\scriptstyle#1$\kern3pt}}
	\savebox{\mysim}{\hbox{$\sim$}}
	\mathbin{\overset{#1}{\kern\z@\resizebox{\wd\mybox}{\ht\mysim}{$\sim$}}}}
\makeatother

\begin{document}
\begin{frontmatter}
\pretitle{Research Article}

\title{Approximate confidence distribution computing}

\begin{aug}
\author[a]{\inits{S.}\fnms{Suzanne} \snm{Thornton}\thanksref{c1}\ead[label=e1]{sthornt1@swarthmore.edu}}
\author[b]{\inits{W.}\fnms{Wentao} \snm{Li}\thanksref{f1}\ead[label=e2]{wentao.li@manchester.ac.uk}}
\author[c]{\inits{M.}\fnms{Minge} \snm{Xie}\thanksref{f1}\ead[label=e3]{mxie@stat.rutgers.edu}}
\thankstext[type=corresp,id=c1]{Corresponding author.}
\address[a]{Address of the First author, \institution{Swarthmore College}, \cny{U.S.A}.\\ \printead{e1}}
\address[b]{Address of the Second author, \institution{The University of Manchester}, \cny{U.K.}.\\ \printead{e2}}
\address[c]{Address of the Third author, \institution{Rutgers, The State University of New Jersey}, \cny{U.S.A}.\\ \printead{e3}} 
\end{aug}


\begin{abstract}
Approximate confidence distribution computing (ACDC) offers a new take on the rapidly developing field of likelihood-free inference from within a frequentist framework. The appeal of this computational method for statistical inference hinges upon the concept of a {\it confidence distribution}, a special type of estimator which is defined with respect to the repeated sampling principle. An ACDC method provides frequentist validation for computational inference in problems with unknown or intractable likelihoods. The main theoretical contribution of this work is the identification of a matching condition necessary for frequentist validity of inference from this method. In addition to providing an example of how a modern understanding of confidence distribution theory can be used to connect Bayesian and frequentist inferential paradigms, we present a case to expand the current scope of so-called approximate Bayesian inference to include non-Bayesian inference by targeting a confidence distribution rather than a posterior. The main practical contribution of this work is the development of a data-driven approach to drive ACDC in both Bayesian or frequentist contexts.
The ACDC algorithm is data-driven by the selection of a data-dependent proposal function, the structure of which is quite general and adaptable to many settings. We explore two numerical examples that both verify the theoretical arguments in the development of ACDC and suggest instances in which ACDC outperform approximate Bayesian computing methods computationally. 
\end{abstract}


\received{\smonth{10} \syear{2022}}
\end{frontmatter}
	
\section{Introduction} \label{sec:intro}

\subsection{Approximate confidence distribution computing}
Approximate confidence distribution computing (ACDC) is a new take on likelihood-free inference within a frequentist setting. The development of this computational method for statistical inference hinges upon the modern notion of a {\it confidence distribution}, a special type of estimator which will be defined shortly. Through targeting this special distribution estimator rather than a specific likelihood or posterior distribution as in variational inference and approximate Bayesian inference, respectively, ACDC provides frequentist validation for inference in complicated settings with an unknown or intractable likelihood where dimension-reducing sufficient summary statistics may not even exist. This work demonstrates another example where confidence distribution estimators connect Bayesian and frequent inference, in the surprising context of computational methods for likelihood-free inference \cite{Xie2013, Thornton2022}.  
	

Let $x_{\rm obs} = \{x_1, \dots, x_n\}$ be an observed sample originating from a data-generating model that belongs to some complex parametric family $M_{\theta}$. Suppose the likelihood function is intractable (either analytically or computationally), but that this model is generative, i.e. given any $\theta \in \P$, we can simulate artificial data from $M_{\theta}$. Let $S_n(\cdot)$ be a summary statistic that maps the sample space into a smaller dimensional space and $r_{n}(\theta)$ be a data-dependent function on the parameter space. The simplest version of ACDC is the rejection algorithm labeled Algorithm 1 below, where $K_{\veps}(u) = \veps^{-1}K(u/ \veps)$ for a kernel function $K(\cdot)$ and $\veps$ is a small positive value, referred to as the {\it tolerance level}.



\noindent
\hrulefill
\vspace{-3mm}

\begin{algorithm}
\caption{Accept-reject approximate confidence distribution computing (ACDC)}\label{alg:rejACC} 
	\begin{tabbing}
	 1. Simulate $\theta_{1},\ldots,\theta_{N}\sim r_{n}(\theta)$; \\
	 For each $i=1,\ldots,N$, \\
	 \quad 2. Simulate $x^{(i)}=\{x_{1}^{(i)},\ldots,x_{n}^{(i)}\}$ from $M_{\theta_i}$;\\ 
	 \quad 3. Accept $\theta_{i}$ with probability $K_{\veps}(s^{(i)}-s_{\rm obs})$,\\ where $s_{{\rm obs}}=S_{n}(x_{\rm obs})$ and $ s^{(i)}=S_{n}(x^{(i)})$.
	\end{tabbing}
\end{algorithm}  

\vspace{-6mm}

\noindent
\hrulefill
\vspace{2mm}

The output of many iterations of Algorithm \ref{alg:rejACC} are potential parameter values, and 
these potential parameter values are draws from the probability density 
\begin{align} 
	q_{\veps}(\theta\mid s_{\rm obs})=
	\frac{\int_{ \cal S}r_{n}(\theta)f_n(s\mid\theta)K_{\veps}(s - s_{\rm obs})\,ds}{\int_{{\cal P} \times{\cal S}}r_{n}(\theta)f_n(s\mid\theta)K_{\veps}(s - s_{\rm obs})\,ds d\theta}.
\label{ACC_dist}
\end{align}
Here, $f_n (  s \mid\theta)$ denotes the likelihood of the summary statistic, implied by the intractable likelihood of the data. Therefore $f_n (  s \mid\theta)$ is typically also intractable. We will refer to $f_n ( s \mid\theta)$ as an {\it s-likelihood} to emphasize that this is distinct from a traditional likelihood function $f_n (x_{obs} \mid\theta)$. We denote the cumulative distribution function corresponding to $q_{\veps}(\theta\mid s_{\rm obs})$ by $Q_{\veps}(\theta\mid s_{\rm obs})$. 

The main contribution of this paper is the establishment of a matching condition under which $Q_{\veps}(\theta\mid s_{\rm obs})$ is
an {\it approximate confidence distribution} for $\theta$ and can be used to derive various types of frequentist inferences. These conditions depend on the choice of $r_{n}(\theta)$ but are rather general and we present a strategy for choosing an appropriate data-dependent function in Section \ref{sec:largeSamp}. 
Practically, this new perspective allows the data to drive the algorithm in a way that can make it  more computationally effective than other existing likelihood-free approaches. Theoretically, this perspective establishes frequentist validation for inference from ACDC based upon general conditions that do not depend on the sufficiency of $s_{obs}$. 

Our justification for these practical and theoretical advantages of ACDC relies on the frequentist notion of a confidence distribution. Some background information on confidence distributions is presented next. To motivate the concept, first consider parameter estimation within a frequentist paradigm. We often desire that our estimators, whether point estimators or interval estimators, have certain properties such as unbiasedness or preform similarly under repeated, randomly sampling. A confidence distribution is an extension of this tradition in that it is a distribution estimate (i.e., it is a sample-dependent distribution function) that satisfies certain desirable properties. Following ~\cite{Xie2013} and ~\cite{Schweder2016}, we define a confidence distribution as follows:
	
{ {\it A sample-dependent function on the parameter space is a {\sc confidence distribution (CD)} for a parameter $\theta$ if 1) For each given sample, the function is a distribution function on the parameter space; 2) The function can provide valid confidence sets of all levels for $\theta$.} } 
	
A confidence distribution has a similar appeal to a Bayesian posterior in that it is a distribution function carrying much information about the parameter. A confidence distribution however, is a frequentist notion which treats the parameter as a fixed, unknown quantity and the sampling of data as the random event. 
A confidence distribution is a sample-dependent function that can be used to estimate the parameter of interest to quantify the uncertainty of the estimation. If $H_n(\cdot)$ is a CD for some parameter $\theta$, then one can simulate $\xi_{CD} \sim H_n{(\cdot)}$, conditional upon the observed data. We will refer to the random estimator $\xi_{CD} \sim H_n{(\cdot)}$ as a {\sc CD-random variable}. 


The function $r_{n}(\theta)$ can be viewed as if it is a data-dependent prior. From a frequentist perspective, the data-dependent function $r_{n}(\theta)$ acts as an initial distribution estimate for $\theta$ and Algorithm~\ref{alg:rejACC} is a way to update this estimate in search of a better-preforming distribution estimate. This is analogous to any updating algorithm in point estimation requiring an initial estimate that is updated in search for a better-performing one (e.g., say, a Newton-Raphson algorithm or an expectation-maximization algorithm). Of critical concern in this perspective is ensuring that the data is not `doubly used' for inference. 
An appropriate choice of the initial distribution estimate, $r_{n}(\theta)$, addresses this concern and a general strategy for choosing $r_n(\theta)$ is proposed later in  
Section \ref{sec:largeSamp}. 
Therefore, this perspective asserts that $Q_{\veps}(\theta\mid s_{\rm obs})$ can be used for valid frequentist inference on $\theta$ (e.g., deriving confidence sets, $p$-values, etc.) 
even if it may sometimes not produce the most efficient estimator (i.e., may not produce the tightest confidence sets for all $\alpha\in (0,1)$ levels). 
	

\subsection{Related work on Approximate Bayesian computing (ABC)}  
Approximate Bayesian computation (ABC)
refers to a family of computing algorithms to approximate posterior densities of $\btheta$ by bypassing direct likelihood evaluations \citep[cf.][]{
Csillery2010,
Cameron2012,
Peters2012}.
The target of an ABC algorithm is the posterior distribution rather than a confidence distribution. 
A simple rejection sampling ABC proceeds in the same manner as Algorithm~\ref{alg:rejACC}, but it replaces $\theta_1, \ldots, \theta_N \sim r_{n}(\theta)$ with $\theta_1, \ldots, \theta_N \sim  \pi(\theta)$, a pre-specified prior distribution for $\theta$, in Step 1. 
In this article, we view ABC as a special case of ACDC where $r_{n}(\theta) = \pi(\theta)$.
The simple rejection sampling ABC is computationally inefficient. Some advanced computing techniques have been used to improve the simple ABC approach. One of such advanced algorithms that is comparable to Algorithm \ref{alg:rejACC} is the  an importance sampling version of ABC described in
Algorithm \ref{alg:ISABC}, where the $r_{n}(\theta)$ from Algorithm \ref{alg:rejACC} is treated as a reference distribution 
to facilitate and improve ABC computing efficiency.

\noindent
\hrulefill

\vspace{-3mm}
\begin{algorithm}
\caption{Importance sampling ABC (IS-ABC)} \label{alg:ISABC}
	\begin{tabbing}
	 1. Simulate $\btheta_{1},\ldots,\btheta_{N}\sim r_{n}(\btheta)$. \\
	 For each $i=1,\ldots,N$, \\
	 \quad 2. Simulate $\bx^{(i)}=\{x_{1}^{(i)},\ldots,x_{n}^{(i)}\}$ from $M_{\btheta}$.\\ 
	 \quad 3. Accept $\btheta_{i}$ with probability $K_{\veps}(s^{(i)}-s_{\rm obs})$,\\ 
	 \quad where $s_{\rm obs}= S_{n}(\bx_{\rm obs})$ and $ s^{(i)}=S_{n}(\bx^{(i)})$, and assign \\ \quad importance weights $w(\theta_i)=\pi(\theta_i)/r_{n}(\theta_i)$.
	\end{tabbing}
\end{algorithm}  

\vspace{-6mm} 

\noindent
\hrulefill

The theoretical argument behind an approximate Bayesian inference (either using the simple rejection sampling ABC or IS-ABC) 
 depends upon $q_{\veps}(\theta\mid s_{\rm obs})$ converging to the posterior, $p(\theta \mid x_{obs}) = {\pi(\theta) f_n ( x_{obs} \mid\theta)}\big/{\int \pi(\theta) f_n ( x_{obs} \mid\theta)
d \theta }$, as the tolerance level approaches zero; c.f., e.g.~\cite{Marin2011} and \cite{Beaumont2019}. However, it is well-known that the quality of this approximation depends not only on the size of $\veps$ (and choice of prior) but, also importantly, upon the choice of summary statistic. If $s_{obs}$ is not sufficient (as is gerenally the case in applications of ABC), then the s-likelihood $f_n (s_{obs} \mid\theta)$ can be very different from the likelihood of the data $f_n ( x_{obs} \mid\theta)$ and thus $q_{\veps}(\theta\mid s_{\rm obs})$ can be a very poor approximation to $p(\theta\mid x_{obs})$, even as $\veps$ approaches zero and $n \rightarrow \infty$.

\begin{figure}
\centering
\caption{{\it The gray curves below represent the target posterior distribution (gray lines), $p(\theta \mid \bx)$, for an $n=100$ IID sample from $Cauchy(\theta=10,0.55)$. The curves in black represent  $q_{\veps}(\theta \mid s_{obs})$ for two different summary statistics, $S_{n_1} = Median(x)$ (left) and $S_{n_2} = \bar{x}$ (right). In each case $\veps = 0.005$.} }\label{fig:ACC}
\includegraphics[
width = .9\linewidth, height=5.2cm,
]{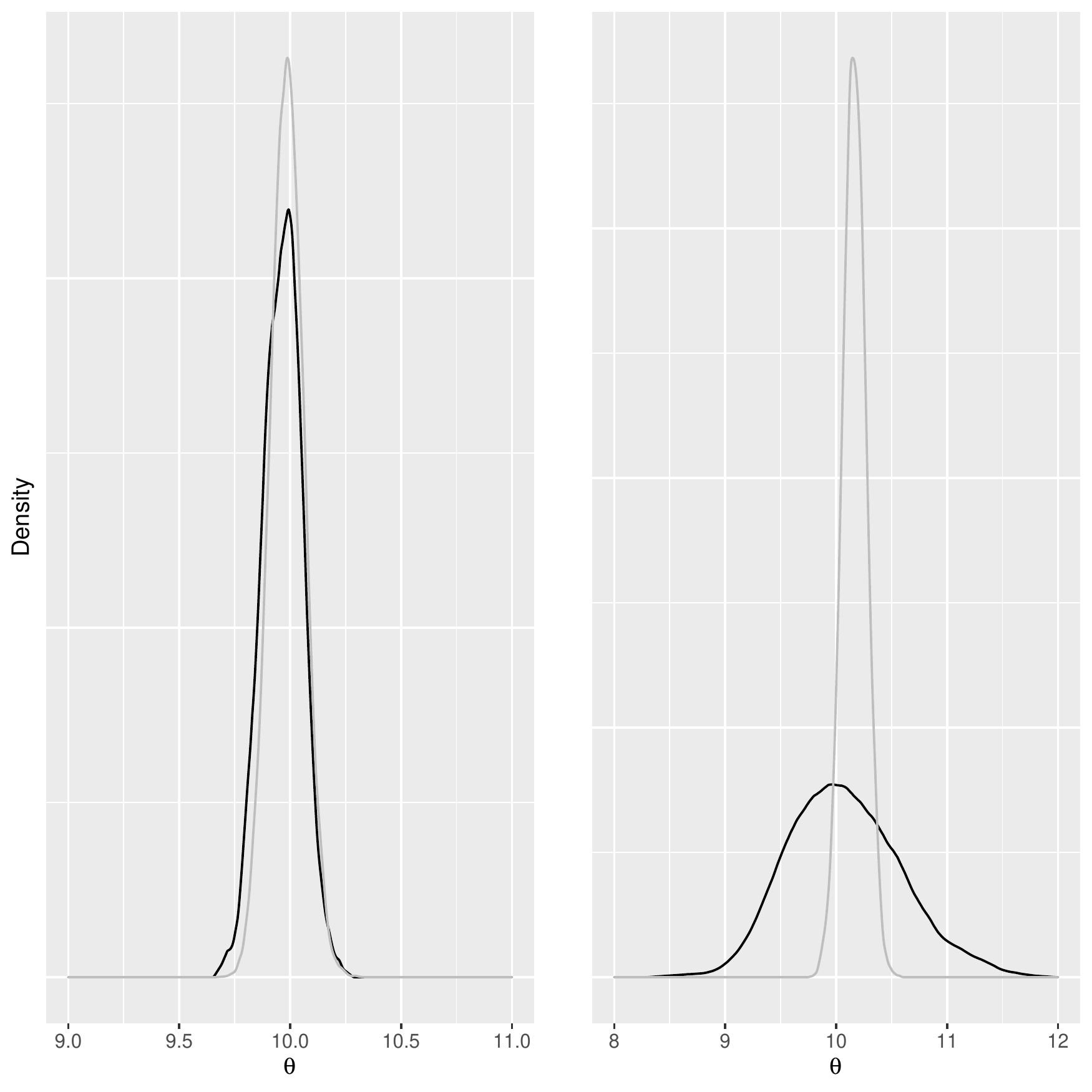}
\end{figure}

For example, consider using Algorithm \ref{alg:rejACC} with two different choices of summary statistic, the sample mean or median, for estimating the location parameter of a random sample ($n=100$) from $Cauchy(10, 0.55)$. If we suppose $r_n(\theta) \propto 1$, then our algorithm is not data-driven and $r_n(\theta)$ instead acts as an uninformative prior. Hence this example corresponds to an accept-reject version of ABC algorithm. 
Figure~\ref{fig:ACC} shows $q_{\veps}(\theta\mid s_{\rm obs})$ for each choice of summary statistic (black lines) where $\veps = 0.005$. 
The posterior distribution (solid gray lines) does not match well with either approximate ABC posterior distribution $q_{\veps}(\theta\mid s_{\rm obs})$ because only the entire data vector itself is sufficient in this example. This example demonstrates how a strictly ABC approach that targets $p(\theta \mid x_{obs})$ can produce inconsistent results and even misleading inferential conclusions.

To quote \cite[]{marquette_statistics_2021}: ``the choice of the summary statistic is essential to ensure ABC produces a reliable approximation to the true posterior distribution." Much of the current literature on ABC methods is appropriately oriented towards the selection and evaluation of the summary statistic. 
The theoretical justification for inference from ACDC on the other hand, does not require an optimal selection of $S_n$. Although less informative summary statistics may lead to less efficient CDs, the validity of the inferential conclusions can remain intact even for less informative (and non-sufficient) summary statistics. (See Sections \ref{sec:main} and \ref{sec:ex}.)


The large sample theoretical results presented in Section \ref{sec:largeSamp} specify conditions under which Algorithm \ref{alg:rejACC} produces an asymptotically normal confidence distribution. These results are similar to those in~\cite{Li2017} 
but our work is distinct because we do not target an approximation to a posterior distribution. Instead, the theoretical results in this section of our paper focus on the properties and performance of ACDC inherited through its connection to CDs. Additionally, in Section \ref{sec:largeSamp} we propose a regression-adjustment technique based on that of \cite{Li2017} and \cite{Blum2010}. This post-processing step for ACDC is applied to Algorithms \ref{alg:rejACC} and \ref{alg:ISABC} in Section \ref{sec:ex} to improve the accuracy of the CDs.


Computationally, the numerical studies in Section~\ref{sec:ex} also suggest that 
accept-reject ACDC 
is more stable than the IS-ABC 
even when both approaches utilize the same data-driven function $r_{n}(\theta)$. This difference in performance is due to the fact that the importance weights  $w(\theta)=\pi(\theta)/r_{n}(\theta)$ in in IS-ABC 
can fluctuate greatly causing numerical instability in the generated parameter values. 
The steep computing cost associated with the generative model is an expansive area of current research on likelihood-free methods including adaptations that decrease the computing cost of approximate Bayesian methods 
such as MCMC methods \cite[]{marjoram2003markov} and sequential Monte Carlo techniques\cite[]{Sisson2007}. Although an exploration of these adaptations is beyond the scope of this paper, we expect that many of these approaches can be readily applied to improve the computational performance of ACDC as well. 
The numerical examples in Section \ref{sec:ex} 
demonstrate how accept-reject ACDC 
accepts more simulations than IS-ABC
suggesting that merely incorporating $r_n(\theta)$ as a data-dependent proposal function is not computationally preferable. 



\subsection{Notation and outline of topics}
In addition to the notation from the introduction, throughout the remainder of paper we will use the following notation. The observed data is $x_{\rm obs} \in \mathscr{X} \subset \mathbb{R}^n$, the summary statistic is a mapping $S_n: \mathscr{X} \rightarrow {\cal S} \subset \mathbb{R}^{d}$ and the observed summary statistic is $s_{\rm obs} = S_n(x_{\rm obs})$. The parameter of interest is $\theta \in \P \subset \mathbb{R}^p$ with $p \leq d < n$; i.e. the number of unknown parameters is no greater than the number of summary statistics and the dimension of the summary statistic is smaller than the dimension of the data. If some function of $S_n$ is an estimator for $\theta$, we will denote this function by $\hat{\theta}_S$. 


The next section presents the core theoretical result of this paper 
which establishes a necessary condition for ACDC methods to produce a valid CD, thereby 
establishing ACDC as a likelihood-free method that provides valid frequentist inference. 
Section \ref{sec:largeSamp} presents general large sample conditions for ACDC that produce asymptotic CDs and establishes precise conditions for an appropriate choice of the data-dependent function $r_n(\theta)$. Section \ref{sec:ex} contains two numerical examples that verify the inferential conclusions of ACDC and illustrate the computational advantages of this data-driven algorithm. Section \ref{sec:discuss} concludes with a brief discussion. All proofs for Sections \ref{sec:thms} and \ref{sec:largeSamp} are contained in the Appendix (and supplementary material).

\section{Establishing frequentist guarantees}\label{sec:thms}
\subsection{General conditions}\label{sec:main} 
In this section, we formally establish conditions under which ACDC 
can be used to produce confidence regions with guaranteed frequentist coverages for any significance level. To motivate our main theoretical result, first consider the simple case of a scalar parameter and a function $\htheta = \hat{\theta}(S_n)$ which maps the summary statistic, $S_n \in {\cal{S}}$, into the parameter space $\cal{P}$. 
	
\noindent {\it Claim: If 
	\begin{equation}
	\label{eq:LR}
	\text{\rm pr}^*(\theta-\htheta \leq t \mid S_n = s_{\rm obs}) = \text{\rm pr}(\htheta-\theta \leq t \mid \theta = \theta_0),
	\end{equation}
then $H_n(t)\stackrel{\hbox{\tiny def}} = 1-Q_{\veps}(2\htheta-t \mid s_{obs})$ is a CD for $\theta$.}

In the claim, $\text{\rm pr}^*(\cdot \mid S_n = s_{\rm obs}) $ refers to the probability measure on the simulation, conditional on the observed summary statistic, and $\text{\rm pr}(\cdot \mid \theta = \theta_{0})$ is the probability measure on the data before it is observed.
The proof of this claim (provided in the appendix) involves showing that $H_n(\theta_0)$ follows a uniform distribution. Once this is established, any $(1-\alpha)100\%$ level confidence interval for $\theta$ can be found by inverting the confidence distribution, $H_n(t)$. 
 
This claim is conceptually similar to the bootstrap central limit theorem which states (conditions under which) the variability of the bootstrap estimator matches the variability induced by the random sampling procedure. Equation \eqref{eq:LR} instead matches the variability induced by the Monte-Carlo sampling  
to the random sampling variability. 
On the left hand side, $\htheta$ is fixed given $s_{\rm obs}$ and the (conditional) probability measure is defined with respect to the Monte-Carlo copies of $\theta$. Thus $\theta$ on the left hand side of this equation plays the role of a CD random variable. On the right hand side, the probability measure is defined with respect to the sampling variability where $\theta$ is the true parameter value. See also  \cite{Thornton2022} for more discussions of similar matchings to link Monte-Carlo randomness with sample randomness across Bayesian, fiducial and frequentist paradigms.


The main condition necessary for valid frequentist inference from ACDC methods is a generalization of the claim above for vector $\theta$. 
	\begin{condition} \label{cond:ACC_interval}
		{\it For $\mathfrak{B}$ a Borel set on $\mathbb{R}^k$,  
		\begin{eqnarray*}
	 \sup_{A\in\mathfrak{B}}\big\| \text{\rm pr}^*\{V(\theta, S_n) \in A \mid  {S_n = s_{\rm obs}} \} \quad\quad\qquad\qquad & \\
		  - \text{\rm pr}\{W( \theta, S_n) \in A \mid {\theta = \theta_0} \}
		\big\| = o_p(\delta_{n,\veps}), &
		\end{eqnarray*} 
		where $\text{\rm pr}^*(\cdot \mid Sn =  s_{\rm obs}) $ refers to the probability measure on the simulation, conditional on the observed summary statistic, and $\text{\rm pr}(\cdot \mid \theta = \theta_{0})$ is the probability measure on the data before it is observed and finally $\delta_{n,\veps}$ is a positive rate of convergence that depends on $n$ and $\veps$. }
	\end{condition}
Rather than consider only the linear functions $(\theta-\htheta)$ and $(\htheta-\theta)$, Condition \ref{cond:ACC_interval} considers any functions $V(\theta, S_n)$ and $W( \theta, S_n)$. (For example, the claim above is a special case of Condition \ref{cond:ACC_interval} where $V(t_1, t_2) = - W(t_1, t_2) = t_1 -\hat{\theta}_{S}(t_2)$.) We use the notation $pr^*\{\cdot \mid S_n = s_{obs}\}$  because this probability measure is defined over a transformation of the $\theta \sim Q_{\veps}(\cdot \mid s_{obs}$).  

Furthermore, Condition \ref{cond:ACC_interval} permits the parameter space and the sample space of the summary statistic to be different from each other. In short, a matching condition on the relationship between two general, multi-dimensional mappings, $V$, $W: \P \times {\cal S}  \rightarrow \mathbb{R}^k$ is the key to establishing when ACDC can be used to produce a confidence distribution for $\theta$.
	

For a given $s_{\rm obs}$ and $\alpha \in (0,1)$,  we can define a set $A_{1 - \alpha} \subset \mathbb{R}^k$  such that, 
	\begin{equation}
	\label{eq:A} 
	\text{pr}^*\{V(\theta, S_n) \in A_{1 - \alpha} \mid S_n = s_{\rm obs}\} = (1 - \alpha) + o(\delta'), 
	\end{equation}  
	where $\delta' > 0$ is a pre-selected small positive precision number, designed to control Monte-Carlo approximation error. If Condition \ref{cond:ACC_interval} holds, then 
	\begin{equation}
	\label{eq:AB}
	\Gamma_{1 - \alpha}(s_{\rm obs}) \stackrel{\hbox{\tiny def}} = \{\theta: W(\theta, s_{\rm obs}) \in A_{1 - \alpha} \} \subset {\cal P}
	\end{equation}
is a level $(1 - \alpha) 100\%$ confidence set for $\theta_0$. We summarize this in the following lemma which is proved in the appendix. 
	
\begin{lemma}\label{main1}
{\it Suppose there exist mappings $V$ and $W: \P \times {\cal S} \rightarrow \mathbb{R}^k$ such that Condition \ref{cond:ACC_interval} holds.
Then, $\text{\rm pr}\{\theta_0 \in \Gamma_{1 - \alpha}(S_n) \mid \theta =  \theta_0 \}  = (1 - \alpha) + o_p(\delta)$, where $\delta = \max\{\delta_{n,\veps},\delta'\}$. If Condition \ref{cond:ACC_interval} holds almost surely, then 
$\text{\rm pr}\{\theta_0 \in \Gamma_{1 - \alpha}(S_n) \mid \theta = \theta_0 \}  \stackrel{\hbox{\tiny a.s.}}= (1 - \alpha) + o(\delta)$.}
\end{lemma}

Nowhere in Lemma \ref{main1} is the sufficiency (or near sufficiency) of $S_{n}$ required. Of course, if the selected summary statistic happens to be sufficient, then inference from the CD with be equivalent to maximum likelihood inference. Furthermore, Lemma \ref{main1} may hold for finite $n$ provided Condition \ref{cond:ACC_interval} does not require $n \rightarrow \infty$, i.e. provided $\delta_{n,\veps}$ only depends on $\veps$. Later in this section we will consider a special case of Lemma \ref{main1} that may be independent of sample-size. 

	
In the next sections we explore some specific situations in which Condition \ref{cond:ACC_interval} holds. First however, we relate equation (\ref{eq:A}) to a random sample from $Q_{\veps}(\cdot \mid s_{\rm obs})$ for vector $\theta$. Suppose $\theta'_{i}$, $i = 1, \ldots, m$, are $m$  draws from $Q_{\veps}(\cdot \mid s_{\rm obs})$ and let ${v}_i =  V(\theta'_{i},  s_{\rm obs})$. The set $A_{1 - \alpha}$ may be a $(1-\alpha)100\%$ contour set of $\{{ v}_1, \ldots, { v}_m\}$ such that $o(\delta') = o(m^{-1/2})$. For example, we can directly use $\{ v_1, \ldots, v_m\}$ to construct a  $100(1-\alpha)\%$ depth contour as $A_{1 - \alpha} = \{\theta : (1/m)\sum_{i=1}^{m}\mathbb{I}{\{\hat D({ v}_i)< \hat D(\theta)\}} \geq \alpha\}$, where $\hat D(\cdot)$ is an empirical depth function on~$\P$ computed from the empirical distribution of $\{{ v}_1, \ldots, { v}_m\}$. See, e.g.,~\cite{Serfling2002} and~\cite{Liu1999} for more on the development of data depth and depth contours in nonparametric multivariate analysis.  
	
\subsection{Finite sample size case} 
We now explore a special case of Lemma \ref{main1} where the mappings $V$ and $W$ correspond an approximate pivot statistic. We call  a mapping $T = T(\theta, S_{n})$ from $\P \times {\cal S} \to \mathbb{R}^{d}$  an {\it approximate pivot statistic}, if
	\begin{equation}\label{eq:apivot}
	\text{pr}\{T(\theta, S_{n}) \in A \mid \theta = \theta_0\} =  \int_{t \in A}
	g(t) d t \,  \{1 + o(\delta^{''})\}, 
	\end{equation}
where $g(t)$ is a density free of $\theta$, $A \subset \mathbb{R}^{d}$ is any Borel set, and $\delta^{''}$ is either zero or a small number (tending to zero) that may or may not depend on the sample size $n$. For example, suppose $S_n | \theta = \lambda \sim \text{Poisson}(\lambda)$. Then, $T( \lambda, S_{n}) = (S_n -  \lambda)/\sqrt{ \lambda}$ is an approximate pivot when $\lambda$ is large, and the density function is $\phi(t) \{1 + o(\lambda^{-1})\}$, where $\phi(t)$ the density function of the standard normal distribution~\cite[]{Cheng1949}. The usual pivotal cases are special examples of approximate pivots that may not rely on large sample theory. Examples of approximate pivots where $\delta^{''}$ is a function of $n$
are discussed later in Section \ref{sec:largeSamp}. 

\noindent
\begin{thm} \label{thm:pivot} 
{\it Suppose $T = T(\theta, S_{n})$ is an approximate pivot statistic that is differentiable with respect to the summary statistic and, for given $t$ and $\btheta$,  let $s_{t, \btheta}$ denote a solution to the equation $t = T(\theta, s)$.  If 
		\begin{equation}\label{eq:req}
		\hbox{$\int_{\cal{P}}  r_{n}(\theta) K_{ \veps}\left(s_{t, \theta}-s_{\rm obs}\right)   d \theta = C$},
		\end{equation}
		where $C$ is a constant free of $t$,
		then, Condition \ref{cond:ACC_interval} holds almost surely, for $V(\theta, S_n) = W(\theta,S_{n}) = T( \theta, S_{n}).$ }
\end{thm} 

A direct implication of Theorem \ref{thm:pivot} is that $\Gamma_{1 - \alpha}(s_{\rm obs})$, as defined in~(\ref{eq:AB}), is a level  $(1 - \alpha)100\%$ confidence region where $\text{pr}\{\theta \in \Gamma_{1 - \alpha}(S_n) \mid  \btheta_0 \}  \stackrel{\hbox{\tiny a.s.}} = (1 - \alpha) + o(\delta)$.
	
The assumption in equation (\ref{eq:req}) needs to be verified on a case-by-case basis. 
Location and scale families contain natural pivot statistics and satisfy these conditions. This is formally stated in the following corollary. 	

\begin{cor}\label{cor:pivot}
{\it 

\noindent (a) Suppose $S_n$ is a point estimator for $\mu$ such that $S_n \sim g_1(S_n - \mu)$ and suppose $r_{n}(\mu) \propto 1$. Then, for any $u$,  
$$|\text{pr}^*(\mu  - S_n  \leq u\mid S_n=s_{obs})-  \text{pr}(S_n  - \mu \leq u \mid \mu= \mu_0)| \stackrel{\hbox{\tiny a.s.}} =  o(1).$$ 

\noindent (b) Suppose $S_n$ is a point estimator for $\sigma$ such that $S_n \sim g_2(S_n/ \sigma )/\sigma$ and suppose $r_{n}(\sigma) \propto 1 / \sigma$. then, for any $v>0$, 
$$\left|\text{pr}^*\left(\frac{\sigma}{S_n}  \leq v \big| S_n = s_{obs}\right)-  \text{pr}\left(\frac{S_n}{\sigma} \leq v \big| \sigma = \sigma_0 \right) \right| \stackrel{\hbox{\tiny a.s.}} =   o(1).$$

\noindent (c) If $S_{n,1}$ and $S_{n,2}$ are point estimators for $\mu$ and $\sigma$, respectively, where $S_{n,1} \sim g_1\{(S_{n,1} - \mu)/ \sigma \}/\sigma $ and $S_{n,2} \sim g_2\left( S_{n,2}/ \sigma\right)/\sigma$ are independent and if $r_{n}(\mu, \sigma) \propto 1 / \sigma$, then, for any $u$ and any $v > 0$, 
\bes 
&\Big|\text{pr}^*\left[\begin{pmatrix}\mu - S_{n,1}  \leq u\\ \frac{\sigma}{S_{n,2}}  \leq v
\end{pmatrix}
\big| 
\begin{pmatrix}
S_{n,1} \\
S_{n,2}
\end{pmatrix} = 
\begin{pmatrix}
s_{1,obs} \\
s_{2, obs}
\end{pmatrix} \right] \qquad \qquad \\
&\qquad \qquad - \text{pr}\left[
\begin{pmatrix}
\mu - S_{n,1}  \leq u \\
\frac{S_{n,2}}{\sigma} \leq v 
\end{pmatrix}
\big| 
\begin{pmatrix}
\mu \\ \sigma 
\end{pmatrix} =
\begin{pmatrix}\mu_0\\  \sigma_0\end{pmatrix}
\right]\Big| \stackrel{\hbox{\tiny a.s.}} =  o(1). 
\ees

\noindent Consequently, $H_1(S_{n,1}, x) = \int_{-\infty}^{x}g_1(S_{n,1}-u)du $ is a CD for $\mu$  
and $H_2(S_{n,2}^2,x) = 1 - \int_{0}^{x} g_2(\hat{\sigma}_{S}/u)du,$ is a CD for $\sigma$. 
}
\end{cor}

Note that Theorem \ref{thm:pivot} and Corollary \ref{cor:pivot} cover some finite sample size scenarios, including the Cauchy example discussed in Section \ref{sec:intro}. For this example, Corollary \ref{cor:pivot} (part (a)) asserts that the different posterior approximations obtained by approximate Bayesian computing with either $S_n = Median(x)$ or $S_n = \bar{x}$ 
are both CDs. 
That is, both densities in black in Figure \ref{fig:cauchy_loc_ex} are densities for confidence distributions of $\theta$. These distribution estimators  lead to valid frequentist inference 
even though neither summary statistic is sufficient. This development represents a departure from the typical asymptotic arguments for likelihood-free computational inference.

This section has considered the case in which the tolerance level, $\veps$, does not necessarily depend on the sample size $n$. In the next section however, the tolerance may depend on the sample size and so we adopt the notation $\veps_n$ to reflect this.

\section{Large sample theory}\label{sec:largeSamp}
\subsection{A Bernstein-von Mises theorem for ACDC}\label{sec:large_n_theory}
In the Bayesian ABC framework, Condition \ref{cond:ACC_interval} holds as $n\rightarrow\infty$ 
by selecting a $\veps_n$ that decreases to zero at a certain rate.
\cite[]{Li2016}. We now verify Condition \ref{cond:ACC_interval} holds more generally for 
ACDC methods that use a  data-dependent $r_{n}(\theta)$, in a large sample setting. The results presented here are generalizations of results in 
\cite{Li2017} and \cite{Li2016}. 
Roughly speaking, the next theorem establishes that the distribution of a centered random draw from $Q_{\veps}(\theta \mid s_{obs})$
and the distribution of its centered expectation (before the data is observed), i.e. $\int \btheta  \, d Q_{\veps}(\theta\mid S_n)$, are asymptotically the same. 

The next condition concerns the asymptotic behavior of the summary statistic is crucial for the proofs of the theorems in this section (see Appendix \ref{Appendix:Thm2and3}). 

\begin{condition} \label{sum_conv}
{\it There exists a sequence $\{a_{n}\}$, satisfying $a_{n}\rightarrow\infty$
		as $n\rightarrow\infty$, a $d$-dimensional vector $s(\btheta)$, a $d\times d$ matrix $A(\btheta)$, and some $\delta_0>0$ such that for $S_{n}\sim f_{n}(\cdot\mid\theta)$
		and all $\btheta\in\mathcal{P}_{0} 
\stackrel{\hbox{\tiny def}} =  \{ \btheta : \|\btheta-\btheta_{0}\| < \delta_0\} \subset \mathcal{P}$, 
		\[
		a_{n}\{\bS_{n}-s(\btheta)\} \stackrel{\hbox{\tiny d}} \rightarrow N\{0,A(\btheta)\},\mbox{ as \ensuremath{n\rightarrow\infty}},
		\]
and $\bs_{\rm obs} \stackrel{\hbox{\tiny P}} \rightarrow  s(\btheta_{0})$.
Furthermore, assume that 

\noindent (i) $s(\btheta), A(\btheta)\in C^{1}(\mathcal{P}_{0})$, $A(\btheta)$ is
		positive definite for all $\theta$; 

\noindent (ii) for any $\delta>0$ there
		exists a $\delta'>0$ such that $\|s(\btheta)-s(\btheta_{0})\|>\delta'$
		for all $\btheta$ such that $\|\btheta-\btheta_{0}\|>\delta$; and
		
\noindent (iii) $I(\theta)  \stackrel{\hbox{\tiny def}} = \left\{ \frac{\partial}{\partial\theta}s(\theta)\right\} ^{T}A(\theta)^{-1}\left\{ \frac{\partial}{\partial\theta}s(\theta)\right\} $
has full rank at $\btheta=\btheta_{0}$.}
\end{condition}

This is a standard condition but, notably, does not depend on the sufficiency of this statistic. Because of this, we refrain from discussing this condition further so we may instead focus on our main contribution, the development of the following regulatory conditions on $r_{n}(\theta)$.  
	
\begin{condition} \label{par_true}
{\it 
For all $\theta \in \mathcal{P}_{0},$ $r_{n}(\theta) \in C^2(\mathcal{P}_{0})$ and $r_{n}(\theta_0)>0$.}
\end{condition}
	
\begin{condition} \label{initial_upper}
{\it There exists a sequence $\{\tau_{n}\}$ 
such that $\tau_{n}=o(a_n)$ and $\sup_{\btheta\in {\cal P}_{0}}\tau_{n}^{-p}r_{n}(\btheta)=O_{p}(1)$.}
\end{condition}
	
\begin{condition} \label{initial_lower}
{\it There exists constants $m$, $M$ such that $0 < m <\mid \tau_{n}^{-p}r_{n}(\btheta_{0})\mid < M < \infty$.}
\end{condition}
	
\begin{condition} \label{initial_gradient}
{\it It holds that $\sup_{\btheta\in\mathbb{R}^{p}}\tau_{n}^{-1} D\{\tau_{n}^{-p}r_{n}(\btheta)\}=O_{p}(1)$.}
\end{condition}
	
Condition \ref{par_true} is a general assumption regarding the differentiability of $r_{n}(\theta)$ within an open neighborhood of the true parameter value. Condition \ref{initial_upper} and \ref{initial_lower} essentially require $r_{n}(\theta)$ to be more dispersed than the s-likelihood within a compact set containing $\theta_0$. They require $r_{n}(\theta)$ converge to a point mass more slowly than $f_{n}(\theta \mid s_{\rm obs})$. Condition \ref{initial_gradient} requires the gradient of the standardized version of $r_{n}(\theta)$ to converge with rate $\tau_n$. These are relatively weak conditions and can be satisfied with locally asymptotic $r_{n}(\btheta)$, for example. Of course, a flat prior used in approximate Bayesian inference
also satisfies Condition \ref{par_true}--\ref{initial_gradient}. The proofs of 
the theorems in this section also require additional conditions (Conditions \ref{kernel_prop}-\ref{cond:likelihood_moments} of Appendix \ref{Appendix:Thm2and3}) that are typical of BvM-type theorems. These additional conditions are not presented here for readability reasons and because they do not directly relate to $r_n(\theta)$ which is our emphasis.  

\begin{thm} \label{thm:ACC_limit_small_bandwidth}
{\it Let $\hat{\theta}_{S} = \hat{\theta}(S_n) =  \int \btheta  \, d Q_{\veps}(\theta\mid S_n).$ Assume $r_{n}(\btheta)$ satisfies Condition \ref{par_true}--\ref{initial_gradient} above and also  Conditions \ref{kernel_prop}--\ref{cond:likelihood_moments} in the supplementary material. If $\veps_{n}=o(a_{n}^{-1})$ 
as $n\rightarrow\infty$, then Condition \ref{cond:ACC_interval} is satisfied with $V(\btheta,S_n)=a_{n}\left(\btheta-\hat{\theta}_{s_{obs}}\right)$ and $W(\btheta,S_{n})=a_{n}\left(\hat{\theta}_{S}-\btheta\right)$.}
\end{thm}
	
Theorem \ref{thm:ACC_limit_small_bandwidth} says when $\veps_{n}=o(a_{n}^{-1})$, the coverage of $\Gamma_{1-\alpha}(s_{\rm obs})$ is asymptotically correct as $n$ and the number of accepted parameter values increase to infinity. In practice, $\hat{\theta}_{S}$ typically will not have a closed form. To construct $\Gamma_{1-\alpha}(s_{\rm obs})$, the value of $\hat{\theta}$ at $S_n=s_{\rm obs}$ can be estimated using the accepted parameter values from ACDC.
Here Condition \ref{cond:ACC_interval} is satisfied by generalizing the limit distributions of the approximate posterior 
in~\cite{Li2017} so they hold also for $Q_{\veps}(\theta\mid s_{obs})$, when $\veps_n=o(a_n^{-1})$. Specifically, for $A$ defined as in equation (\ref{eq:A}),
\begin{eqnarray}
&\sup_{A\in\mathfrak{B}^{p}}\Big| \int_{\{\btheta: \, a_{n}(\btheta-\hat{\theta})\in A\}} d Q_{\veps}(\btheta \mid   \bs_{{\rm obs}})-\qquad \qquad \notag \\
&\qquad \qquad \qquad \int_{A}N\{t;0,I(\btheta_{0})^{-1}\}\,dt \Big| \stackrel{\hbox{\tiny P}} \rightarrow 0 
\end{eqnarray}\label{thm2_uncertainty}
and 
\begin{equation}
a_{n}(\hat{\theta}-\btheta_{0}) \stackrel{\hbox{\tiny d}} \rightarrow N\{0,I(\btheta_{0})^{-1}\}, \label{thm2_ptestimate}
\end{equation}
as $n\rightarrow\infty$, where $I(\theta_0)$ is a non-singular matrix defined in Condition \ref{sum_conv}. Thus inference based on $Q_{\veps}(\theta \mid s_{obs})$ is valid for $n \rightarrow \infty$ regardless of whether or not $r_{n}(\theta)$ depends on the data. For the same tolerance level, Theorem \ref{thm:ACC_limit_small_bandwidth} asserts that the limiting distribution of $Q_{\veps}(\theta \mid S_n)$ matches the limiting distribution of the approximate posterior from \cite{Li2017} which is the output distribution of the accept-reject version of ABC. 
In comparison however, Algorithm \ref{alg:rejACC} has a better acceptance rate since the data-dependent $r_{n}(\theta)$ will concentrate more probability mass around $\theta_0$ than a typical prior. 

Although inference from ACDC is validated with $\veps_n=o(a_n^{-1})$, a well-known issue in approximate Bayesian literature is that this tolerance level is too small in practice, causing the acceptance rate to degenerate 
as $n \rightarrow \infty$ for any proposal distribution~\cite[]{Li2016}. Obviously ACDC methods will suffer from this same issue. (For an example with Normal data, see Appendix \ref{sec:degen_of_AR}.)

One remedy that relaxes the restriction on $\veps_n$  
is to post-process the sample from $Q_{\veps}(\theta \mid s_{obs})$ with a regression adjustment\cite[]{beaumont2002}. When the data-generating model is correctly specified, the regression adjusted sample correctly quantifies the CD uncertainty 
and yields an accurate point estimate with $\veps_n$ decaying at a rate of $o(a_{n}^{-3/5})$\cite[]{Li2017}. 
	
Let $\btheta^{*}=\btheta-\beta_{\veps}(\bs-\bs_{{\rm obs}})$ be the post-processed sample from $Q_{\veps}(\theta \mid s_{obs})$, where $\beta_{\veps}$ is the minimizer from 
\begin{equation*}
 \begin{pmatrix}\alpha_{\veps}\\ \beta_{\veps}\end{pmatrix}=
\underset{\alpha\in\mathbb{R}^p,\beta\in\mathbb{R}^{d\times p}}{\arg\min} 
E_{\veps}\left\{\|\btheta-\alpha-\beta(\bs-\bs_{{\rm obs}})\|^{2}\mid \bs_{{\rm obs}}\right\} 
\end{equation*}
for expectation under the joint distribution of accepted $\theta$ values and corresponding summary statistics. 

\begin{thm}\label{thm:ACC_limit_large_bandwidth}
{\it Under the conditions of Theorem \ref{thm:ACC_limit_small_bandwidth}, 
if $\veps_{n}=o\left(a_{n}^{-3/5}\right)$ as $n\rightarrow\infty$, Condition \ref{cond:ACC_interval} holds with 
$V(\btheta,S_n)=a_{n}(\btheta^{*}-\hat{\theta}_{s_{obs}}^*)$ and 
$W(\btheta,S_{n})=a_{n}(\hat{\theta}_{S}^*-\btheta)$, 
where $\hat{\theta}_{S}^{*}$ is the expectation of the post-processed observations of the CD random variable.  }
\end{thm}
		
Here, Condition \ref{cond:ACC_interval} is implied by the following convergence results (where $A$ defined as in equation (\ref{eq:A})),
\begin{eqnarray*}
&\sup_{A\in\mathfrak{B}^{p}}\Big| \int_{\{\btheta: \, a_{n}(\btheta -\hat{\theta}^{*}) \in A\}} d Q_{\veps}^*(\btheta \mid   \bs_{{\rm obs}}) - \qquad \qquad\\ 
&\qquad \qquad\qquad \qquad\int_{A}N\{t;0,I(\btheta_{0})^{-1}\}\,dt\Big|
\stackrel{\hbox{\tiny P}} \rightarrow 0,
\end{eqnarray*}
and 
\[a_{n}(\hat{\theta}_{S}^{*}-\btheta_{0}) \stackrel{\hbox{\tiny d}} \rightarrow  N\{0,I(\btheta_{0})^{-1}\}, \]
as $n\rightarrow\infty$. The limiting distributions above are the same as those in \eqref{thm2_uncertainty} and \eqref{thm2_ptestimate}, therefore $\Gamma_{1-\alpha}(\boldsymbol{s}_{\rm obs})$ constructed using the post-processed sample achieve the same efficiency as those using original ACDC sample of $\theta$ values. The benefit of permitting larger 
tolerance levels is a huge improvement in the computing costs associated with ACDC.

\subsection{Designing $r_{n}$}\label{sec:rn}
Condition \ref{initial_upper} implies that in practice, one must take care to choose $r_{n}(\theta)$ so that its growth with respect to the sample size is slower than the growth of the s-likelihood. In this section we propose a generic algorithm to construct such an $r_{n}(\theta)$ based on sub-setting the observed data. 

Notably, there is a trade-off in ACDC inference between faster computations and guaranteed coverage of the approximate CD based confidence intervals (or regions). When $r_{n}(\theta)$ grows at a similar rate as the s-likelihood for $n \to \infty$, the computing time for ACDC methods may be reduced but this risks violating Conditions \ref{initial_upper}--\ref{initial_gradient}. If these assumptions are violated, the resulting simulations do not necessarily form a CD and consequently, inference 
may not be valid in terms of producing confidence sets with guaranteed coverage. Therefore, $r_{n}(\theta)$ should be designed such that its convergence rate is bounded away from that of the {\it s-}likelihood. The minibatch scheme presented below is one way to ensure
$r_n(\theta)$ is approriately bounded. 

Assume that a point estimator $\htheta(z)$ of $\theta$ can be computed for a dataset, $z$, of any size. 
	
\begin{description}
\item [\textbf{Minibatch scheme}]  
\end{description}
\begin{enumerate}
	\item Choose $k$ subsets of the observations, each with size $n^{\nu}$ for some $0<\nu<1$. 
	\item For each subset $z_{i}$ of $x_{\rm obs}$, compute the point estimate $\hat{\theta}_{\text{S},i}=\htheta(z_{i})$, for $i=1,\ldots,k$. 
	\item Let $r_{n}(\theta)=(1/ kh) \sum_{i=1}^{k}K\left\{h^{-1}\|\theta-\hat{\theta}_{\text{S},i}\|\right\},$ where $h>0$ is the bandwidth of the kernel density  estimate using $\{\hat{\theta}_{\text{S},1},\ldots,\hat{\theta}_{\text{S},k}\}$ and kernel function $K$.
\end{enumerate}

If $\htheta$ is consistent, then 
for $\nu<3/5$, $r_{n}(\theta)$ as obtained by this minibatch procedure will satisfy Conditions \ref{initial_upper}--\ref{initial_gradient}. Based on our experience, if $n$ is large one may simply choose $\nu =1/2$ to partition the data. For small $n$, say $n<100$, it is better to select $\nu>1/2$ and to overlap the subsets (or ``mini" batches of the observed data) so that each subset contains a reasonable number of observations. For a given  summary statistic, there are many methods to construct this type of point estimator including:
a minimum distance-based optimizer~\cite[]{Gourieroux1993,mcfadden1989method}, the synthetic likelihood method and its variants~\cite[]{wood2010statistical,fasiolo2018extended}, or accept-reject ACDC 
with $\hat{\theta}_{S} = E\{\theta\mid S_{n}(z_i)\}$, the s-likelihood-based expectation over a subset of the observed data. The choice of $\htheta$ does not need to be an accurate estimator 
since it is only used to construct the initial rough estimate of a CD for $\theta$. But a heavily biased $\htheta$ causes biases in confidence sets derived from the CD, since 
$r_n(\theta)$ does not cover parameter values resulting in high values of $f_n(s\mid\theta)$ very well. In practice, the computing cost 
will depend on which particular optimization scheme is followed. However, a full study on the selection of $\hat{\btheta}_S$ is beyond the scope of this paper. 
	
The computational cost associated with implementing  the minibatch scheme is comparable to 
the cost of constructing a proposal distribution for IS-ABC methods.
Multiple runs to compute $\hat{\theta}_{\text{S},i}$ values can be parallelized easily and any procedure to obtain a proposal distribution for IS-ABC 
can be applied on the mini batches of data to yield a point estimate for $\theta$. For example, for each subset $z_i$, the conditional mean $E\{\theta\mid S_{n}(z_i)\}$ can be estimated by population Monte Carlo ABC 
on $S_{n}(z_i)$. This is not any more computationally expensive than computing the same estimate on the full data. This, together with the fact that accept-reject ACDC accepts more simulations than IS-ABC, 
make ACDC the favorable choice 
in terms of overall computational performance. The numerical examples in Section \ref{sec:ex} support this conclusion.    
	
At this point, our reader may wonder if $\htheta,$ can be computed, why not simply use a non-parametric bootstrap method to construct confidence sets? Although it requires no likelihood evaluation, this method has two significant drawbacks. First, the non-parametric bootstrap method is heavily affected by the quality of $\htheta$. For example, a bootstrapped confidence interval for $\theta$ is based on quantiles of $\htheta$ from simulated data. A poor estimator typically leads to poor performing confidence sets. In contrast, in ACDC methods, $\htheta$ is only used to construct the initial distribution estimate which is then updated by the data. Second, when it is more computationally expensive to obtain $\htheta$ than the summary statistic, the non-parametric bootstrap will be much more costly than ACDC methods since $\htheta$ must be calculated for each pseudo data set. Example \ref{sec:ricker} in the next section illustrates such an example.

\section{Numerical examples} \label{sec:ex}

\subsection{Location and scale parameters for Cauchy data} \label{sec:cauchy}

In the Cauchy example presented in Figure \ref{fig:ACC} we saw how the lack of a sufficient summary statistic can change the validity of inferential conclusions from an approximate Bayesian inference approach. 
Through a CD perspective however, the inferential conclusions from ACDC 
are valid under the frequentist criterion even if the summary statistic is not sufficient. Provided Condition 1 is satisfied, different summary statistics produce different CDs. Here we present a continuation of this Cauchy example where random data ($n=400$) is drawn from a $Cauchy(\theta, \tau)$ distribution with data-generating parameter values $(\theta_0,\tau_0)=(10,0.55)$. We investigate the performance of $500$ independent $95\%$ confidence intervals for $\theta$ alone (settings one and two) and $\tau$ alone (setting three) and $500$ independent $95\%$ confidence regions when both parameters are unknown (settings four and five).


In each setting, the confidence intervals (regions) are generated for both accept-reject ACDC and IS-ABC utilizing the same 
minibatch scheme to construct $r_n$ with the median and/or the median absolute deviation (MAD) as point estimators and $v=1/2$. The main difference in the two algorithms is the use of $r_n$. In former 
$r_n$ is a data-driven initial CD estimate whereas 
the latter represents a Bayesian approach that assumes an uninformative prior on the parameter space and employs $r_n(\theta)$ as the proposal distribution for the importance sampling updates. Both algorithms are improved by adapting the regression adjustments mentioned in Section \ref{sec:large_n_theory}, so the output for every run of each algorithm is post-processed in this manner. 

Table \ref{EX1_results} compares frequentist coverage proportions of confidence regions from both algorithms. The acceptance proportion determines how many simulated parameter values are kept and thus is directly related to the tolerance level. Most coverage rates are close to the nominal levels when the acceptance proportion is small, which is expected from the asymptotic theory in Section 3. Overall the coverage performance is similar for both algorithms. For settings one, three, and five with informative summary statistics, both algorithms give similar confidence regions which undercover a bit in the finite-sample regime. For settings two and four with less informative summary statistics, accept-reject ACDC 
is preferable because it produces tighter confidence bounds. 

\begin{table}
\caption{{\it Coverage proportions of confidence sets from ACDC applied to Cauchy data under five different settings. Coverage is calculated over $500$ independent runs that draw a $n=400$ IID sample from a $Cauchy(\theta = 10, \tau = 0.55)$ distribution. The Monte Carlo sample size for both 
algorithms is $50,000$ and the nominal coverage level in every setting is $95\%$. The last column displays the median ratio of the sizes of confidence sets from accept-reject ACDC 
divided by those from IS-ABC. }} 
\resizebox{0.5\textwidth}{!}{
\begin{tabular}{lllll}
\hline
\textbf{}                               & \textbf{Acceptance} & \textbf{ACDC}    & \textbf{IS-ABC}  & \textbf{Ratio of Widths/}     \\
                                        & \textbf{proportion} &  \textbf{Coverage}    &   \textbf{Coverage}   & \textbf{Volumes}              \\ \hline
\multicolumn{5}{l}{\textbf{Setting 1: $\theta$ unknown}}         \\ \hline
$S_n = Median(x)$                                                        & $0.005$             & $0.93$             & $0.94$         & $0.94$               \\
                                                                         & $0.05$              & $0.94$             & $0.94$        & $0.94$                \\
                                                                         & $0.10$              & $0.93$             & $0.94$        & $0.94$                \\ \hline
\multicolumn{5}{l}{\textbf{Setting 2: $\theta$ unknown}}                                                                                                                                          \\ \hline
$S_n = \bar{x}$                                                          & $0.005$             & $0.97$             & $0.98$         & $0.65$               \\
                                                                         & $0.05$              & $0.97$             & $0.98$        & $0.60$                \\
                                                                         & $0.10$              & $0.97$             & $0.98$        & $0.56$                \\   \hline
\multicolumn{5}{l}{\textbf{Setting 3: $\tau$ unknown}}                                                                      \\ \hline
$S_n = MAD(x)$                                                           & $0.005$             & $0.93$             & $0.94$         & $1.00$               \\
                                                                         & $0.05$              & $0.92$             & $0.94$        & $1.00$                \\
                                                                         & $0.10$              & $0.93$             & $0.94$        & $1.00$                \\ \hline
\multicolumn{5}{l}{\textbf{Setting 4: $(\theta, \tau)'$ both unknown}}               \\ \hline
\multirow{2}{*}{$S_n = \begin{pmatrix} \bar{x} \\ SD(x)\end{pmatrix}$}   & $0.005$             & $0.96$             & $0.96$         & $0.58$               \\
                                                                         & $0.05$              & $0.99$             & $0.98$        & $0.48$                \\
                                                                         & $0.10$              & $0.99$             & $0.98$        & $0.47$                \\ \hline
\multicolumn{5}{l}{\textbf{Setting 5: $(\theta, \tau)'$ both unknown}}          \\ \hline
\multirow{2}{*}{$S_n = \begin{pmatrix}Median(x) \\ MAD(x)\end{pmatrix}$} & $0.005$             & $0.91$             & $0.93$         & $0.98$               \\
                                                                         & $0.05$              & $0.94$             & $0.96$        & $1.00$                \\
                                                                         & $0.10$              & $0.94$             & $0.97$        & $1.00$                \\     
\end{tabular}\label{EX1_results}}
\end{table}

\begin{figure}
\centering
\caption{{\it These are densities of point estimators from accept-reject ACDC (red) and IS-ABC (black) for the $500$ independent data sets for each of the five settings in Table \ref{EX1_results}. Additionally, this figure shows a box plot of the relative sizes of the $500$ confidence sets, that is, the length (or volume) of regions produced by accept-reject ACDC 
divided by those of IS-ABC.}}\label{fig:cauchy_loc_ex}
\includegraphics[width=0.9\linewidth]{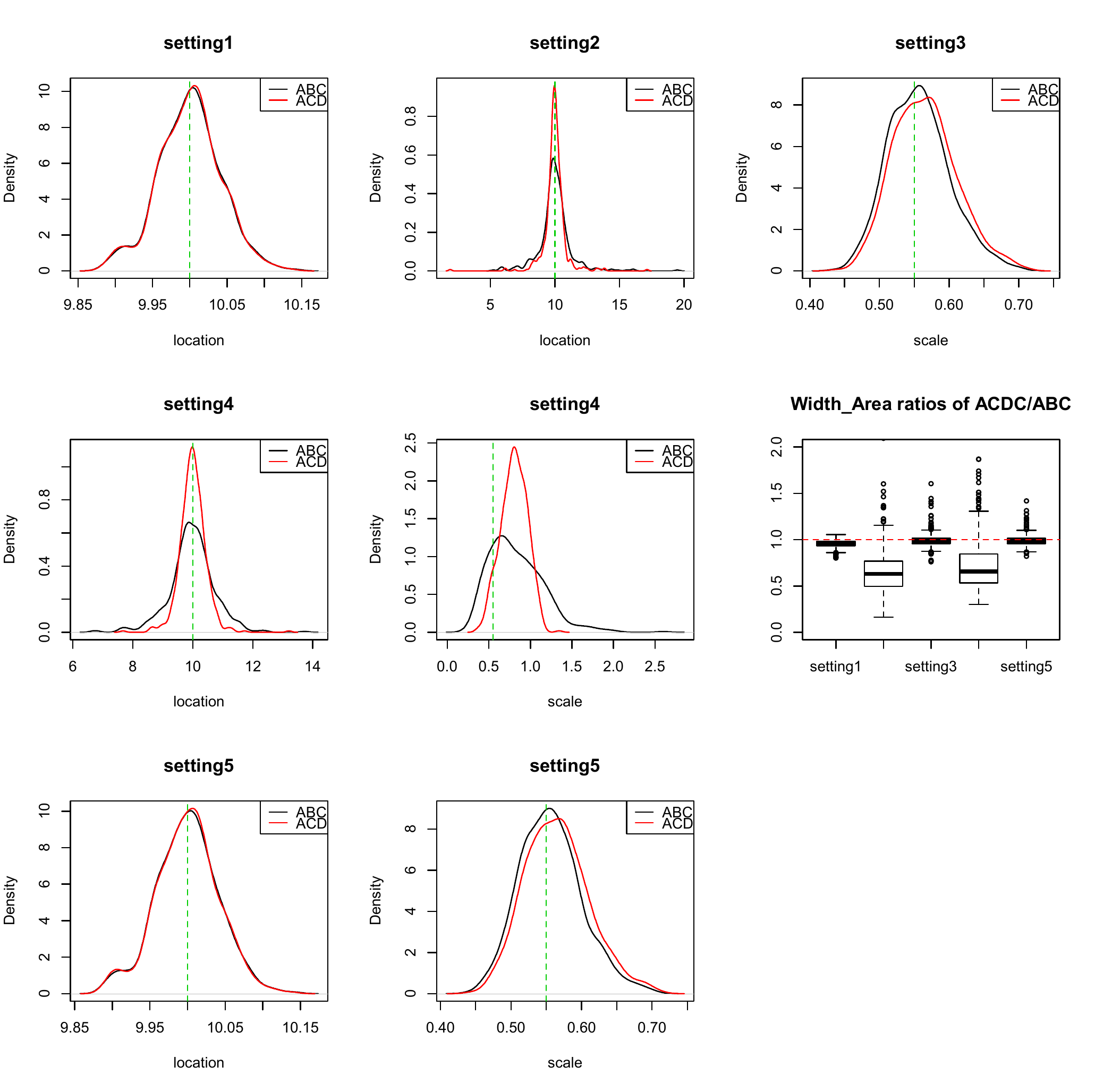}
\end{figure}

The main reason for the favorable performance of accept-reject ACDC 
is related to the skewed importance weights for IS-ABC. This can be seen in the sizes of the confidence sets of the two algorithms in Table \ref{EX1_results} but is even more clear when comparing the CD densities of each method as in Figure \ref{fig:cauchy_loc_ex}. Figure \ref{fig:cauchy_loc_ex} shows the impact of importance weights in IS-ABC 
on the variances of point estimators and CDs. 
For settings 1,3 and 5 where an informative summary statistic is used, the importance weights do not much affect either the point estimator or resulting CDs. In these cases, $r_n(\theta)$ is a good proposal distribution according to the criteria in \cite{Li2016}. For settings 2 and 4 however, where the summary statistic is less informative, Figure \ref{fig:cauchy_loc_ex} shows how the importance weights inflate both point estimate and CD variances with Monte Carlo variation in IS-ABC. One reason for the severe skewedness in the importance weights $\pi(\theta)/r_n(\theta)$ is that the high variance of $S_n$ means more parameter simulations are accepted in the tails of $r_n(\theta)$. This results in broader confidence regions for IS-ABC than accept-reject ACDC.

This numerical study validates inference for both 
accept-reject ACDC and IS-ABC 
even in the case where typical asymptotic arguments do not apply (settings 2 and 4). 
Furthermore, this example demonstrates two valid but distinct uses of the minibatch scheme for constructing a data driven distribution estimator. In Algorithm \ref{alg:rejACC}, $r_n$ drives the search for a distribution estimator or acts as a data-dependent prior within a Bayesian context. In Algorithm \ref{alg:ISABC}, $r_n$ acts as a proposal distribution for ABC with a flat prior. The computational differences in the performance of confidence regions in this example suggest that the former application of $r_n$ is preferable to the latter if the summary statistic is not very informative. Interestingly, even though the Bayesian IS-ABC algorithm fails to give us Bayesian posterior distributions, it can still provide us valid frequentist inference.

\subsection{Mulit-parameter inference for a Ricker model}\label{sec:ricker}
A Ricker map is a non-linear dynamical system, often used in Ecology, that describes how a population changes over time. The population, $N_t$, is noisily observed and is described by the following model,
\begin{align*}
	& y_t \sim \mbox{Pois}(\phi N_t),\\
	& N_t = r N_{t-1}e^{-N_{t-1}+e_t}, e_t\sim N(0,\sigma^2),
\end{align*}
where $t=1,\ldots,T$ and parameters $r$, $\phi$ and $\sigma$ are positive constants, interpreted as the intrinsic growth rate of the population, a scale parameter, and the environmental noise, respectively. This model is computationally challenging since its likelihood function is intractable for $\sigma > 0$ and is highly irregular in certain regions of the parameter space. 

We investigate the performance of confidence bounds for each parameter marginally and two pairs of parameters jointly. We follow the setting and the choice of summary statistics in \cite{wood2010statistical}. The output of both algorithms are post-processed using the regression adjustment.  

In the minibatch scheme, for the point estimator we use $E\{\theta\mid S_n(z_i)\}$ estimated by the population Monte Carlo version of IS-ABC. 
The maximum synthetic likelihood estimator proposed in \cite{wood2010statistical} was also tried, but the estimates obtained this way over-concentrated in a certain area of the parameter space. The 
corresponding $r_n(\theta)$ did not cover the target mass very well, causing biases in the coverage levels. Instead, $r_n(\theta)$ is used to initialise the population Monte Carlo iterations. Since the sample size is small in this example, overlapping minibatches are chosen with a total number of $40$ where each minibatch contains a series of  length $10$. In this example, the parametric bootstrap method is not computationally feasible because it is computationally expensive to use the simulation-based methods in obtaining the point estimates.

\begin{figure}
\centering
\caption{{\it These are densities of point estimators from accept-reject ACDC (red) and IS-ABC (black) for the $150$ independent data sets produced by the Ricker model. Additionally, this figure shows a box plot of the ratio of the sizes of the $150$ confidence sets, that is, the length (or volume) of regions produced by accept-reject ACDC divided by those of IS-ABC.
}}\label{fig:EX2_results}
\includegraphics[width=0.9\linewidth]{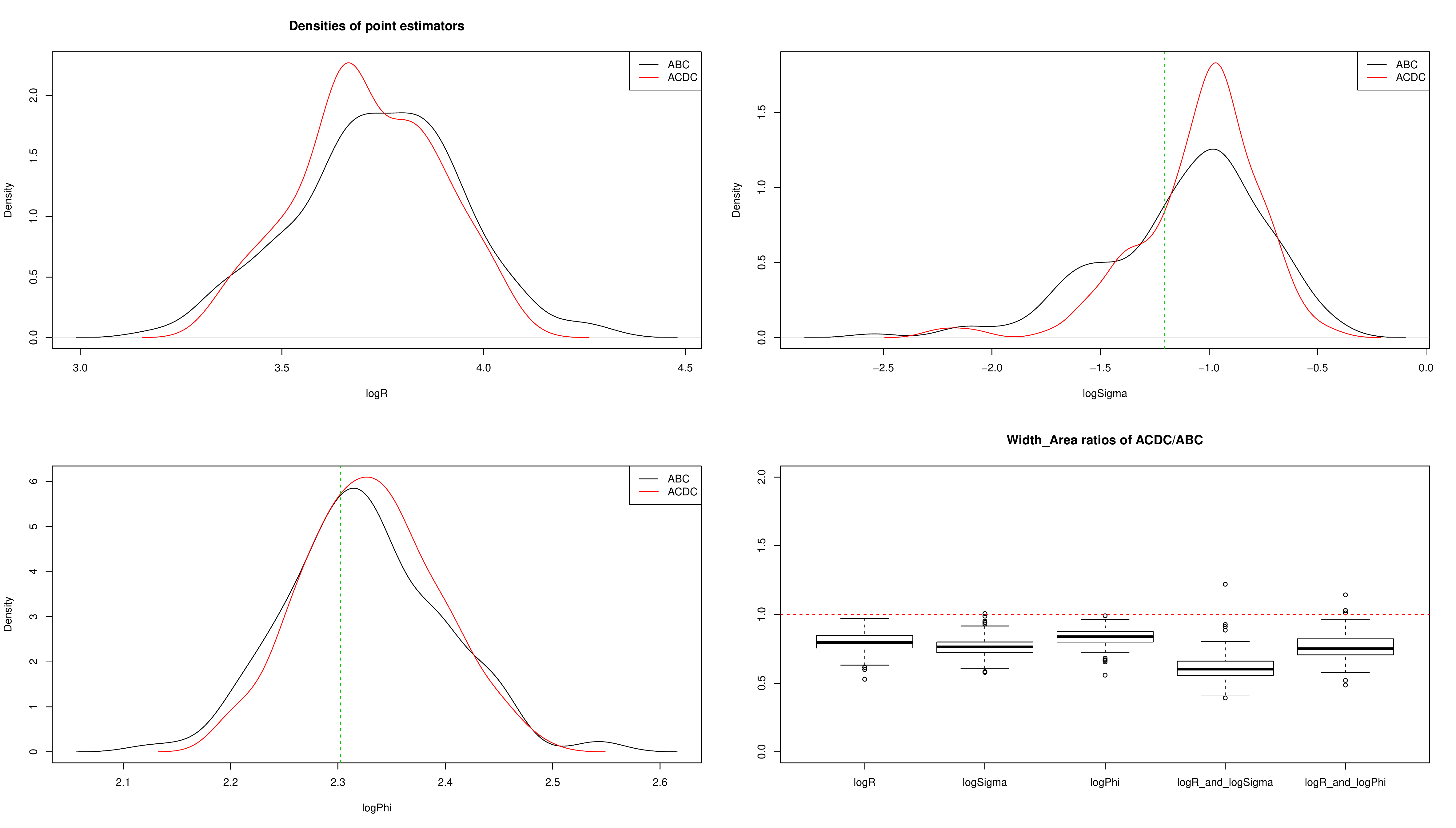}
\end{figure}

In Table \ref{EX2_results}, when the acceptance proportion is small, most coverage rates of 
accept-reject ACDC are close to the nominal level. In contrast, the  confidence bounds from IS-ABC 
display more over-coverage, indicating an even  smaller $\varepsilon$ is needed to reduce the variance inflation. Furthermore, all ACDC confidence regions 
are tighter with a size reduction up to $51\%$. The box plots in Figure \ref{fig:EX2_results} show that the CD variances from IS-ABC are inflated substantially by the importance weights, resulting in broader confidence regions as observed in the last column of Table \ref{EX2_results}.

\begin{table}
\caption{{\it Coverage proportions of marginal confidence intervals (or joint confidence regions) for accept-reject ACDC and IS-ABC applied to Ricker data. Coverage is calculated over $150$ independent runs that produce observations from $t=51$ to $100$ for data generated by a Ricker model with $(r,\sigma, \phi)=(e^{3.8},0.3,10)$. The Monte Carlo sample size for both algorithms is $50,000$ and the nominal coverage level in every setting is $95\%$. The last column displays the median ratio of the sizes of confidence sets from accept-reject ACDC divided by those from IS-ABC.}}
\resizebox{0.5\textwidth}{!}{
\begin{tabular}{lllll}
\hline
\textbf{}                               & \textbf{Acceptance} & \textbf{ACDC}    & \textbf{IS-ABC}  & \textbf{Ratio of Widths/}     \\
                                        & \textbf{proportion} &  \textbf{Coverage}    &   \textbf{Coverage}   & \textbf{Volumes}              \\ \hline
\multicolumn{5}{l}{\textbf{Setting 1: $log(r)$ unknown}}         \\ \hline
           & $0.005$            & $0.953$             & $0.980$       & $0.793$               \\
          & $0.05$              & $0.960$             & $0.987$        & $0.734$                \\
          & $0.10$              & $0.967$             & $0.987$        & $0.707$                \\ \hline
\multicolumn{5}{l}{\textbf{Setting 2: $log(\sigma)$ unknown}}              \\ \hline
          & $0.005$             & $0.967$             & $0.987$         & $0.782$               \\
          & $0.05$              & $0.987$             & $0.993$        & $0.732$                \\
          & $0.10$              & $0.987$             & $0.993$        & $0.717$                \\ \hline
\multicolumn{5}{l}{\textbf{Setting 3: $log(\phi)$ unknown}}               \\ \hline
          & $0.005$             & $0.953$             & $0.967$         & $0.828$               \\
          & $0.05$              & $0.947$             & $0.993$        & $0.762$                \\
          & $0.10$              & $0.960$             & $0.987$        & $0.734$                \\ \hline
\multicolumn{5}{l}{\textbf{Setting 4: $(log(r), log(\sigma))'$} unknown}               \\ \hline
          & $0.005$             & $0.960$             & $0.947$         & $0.611$               \\
          & $0.05$              & $0.960$             & $0.973$        & $0.519$                \\
          & $0.10$              & $0.960$             & $0.947$        & $0.484$                \\ \hline
\multicolumn{5}{l}{\textbf{Setting 5: $(log(r), log(\phi))'$ unknown}}          \\ \hline
          & $0.005$             & $0.973$             & $0.987$         & $0.749$               \\
          & $0.05$              & $1.0$             & $1.0$        & $0.619$                \\
          & $0.10$              & $1.0$             & $1.0$        & $0.557$                
\end{tabular}\label{EX2_results}}
\end{table}

As in Section \ref{sec:cauchy}, this numerical example validates the inferential conclusions from both algorithms but here we see accept-reject ACDC 
consistently producing tighter confidence regions than IS-ABC. The summary statistics in this example were carefully selected to be informative based on domain knowledge. Nevertheless, accept-reject ACDC can still avoid the excessive Monte Carlo variation that impedes IS-ABC.

\section{Discussion} \label{sec:discuss}


In this article, we propose ACDC as a new inference-based approach to likelihood-free methods. It can provide valid frequestist inference for target parameters from data without a tractable likelihood.
ACDC can be viewed as an extension of ABC but, crucially, ACDC does not require any Bayesian assumptions nor does the validity of inferential conclusions depend upon the near-sufficiency of the summary statistic. Computationally, an ACDC approach 
is preferable when compared to the corresponding IS-ABC method which suffers from skewed importance weights.

The main theoretical contribution of this work 
is the identification of a matching condition (Condition \ref{main1}) necessary for valid frequentist inference from ACDC methods. This condition is similar to the theoretical support for Bootstrap estimation and is met in cases that rely on typical asymptotic arguments (e.g. reference citations in Section \ref{sec:largeSamp}) but also applies to certain small-sample cases.
Additionally, a key practical contribution of this work 
is the general minibatch method for initializing ACDC estimators.
This approach guides the search for a well-behaved distribution estimator using a data-dependent distribution $r_{n}(\theta)$. This can result in improved computational performance even compared to an IS-ABC method that is similarly data-driven.
In cases where $r_{n}(\theta)$ does not yield reasonable acceptance probabilities 
we expect that many of the established techniques used in ABC can be readily adapted to ACDC to further improve its computational performance without sacrificing the frequentist inferential guarantees.

An ACDC approach quantifies the uncertainty in estimation by drawing upon a direct connection to confidence distribution estimators. Different choices of summary statistic yield different approximate CDs, some producing tighter confidence sets than others. However, inference from ACDC is validated, regardless of the sufficiency of $S_n$, provided Condition \ref{cond:ACC_interval} can be established. Within a Bayesian framework, there is no clear way to choose among different posterior approximations associated with 
different summary statistics. By pivoting to a frequentist perspective, different summary statistics produce different (CD) estimators but all of these estimators are well-behaved in the long run, yielding  valid inferential statements about $\theta$. 
Supported by the theoretical developments and examples in this paper, it appears as though ACDC provides a more parsimonious solution to validating  likelihood-free inference than attempts to reconcile differences among posteriors and their various approximations. 

\section*{Acknowledgment}
The research is supported in part by research grants from the US National Science Foundation 
(DMS1812048, DMS2015373 and DMS2027855). 
This research stems from a chapter of the first author's PhD Thesis. 
The first author also acknowledges the generous graduate support from Rutgers University.
	
\appendix

\section{Claim in Section 2} 
\begin{proof} First note that $H_n(t) = 1 - Q_{\veps}(2\hat{\theta}-t \mid S_n = s_{obs})$ is a sample-dependent cumulative distribution function on the parameter space. 
By equation (\ref{eq:LR}), we denote both sides as $G(t)$, i.e. $G(t) = \text{pr}^*\{ \theta - \hat \theta_S \leq t \mid S_n = s_{obs} \} = \text{pr}\{\hat \theta_S - \theta \leq t \mid  \theta = \theta_0\}$.
Now we can write $H_n(\theta_0)  
		= \text{pr}^*(2\htheta - \theta \leq \theta_0 \mid  S_n = s_{\rm obs} )
		= \text{pr}^*(\theta - \htheta \geq \htheta- \theta_0 \mid  S_n = s_{\rm obs} )
		= 1 - G(\htheta - \theta_0)$, for $G(t) = \text{pr}(\htheta - \theta \leq t \mid \theta = \theta_0)$. The last equality holds by equation \eqref{eq:LR}. Since $G(\htheta - \theta_0) \mid  \theta_0 \sim Unif(0,1)$ with respect to the sampling variability of $\htheta$, $H_n(\theta_0) = H_n(\theta_0, s_{\rm obs}) \sim Unif(0,1)$. By definition, $H_n(\cdot)$ is a confidence distribution for $\theta$.  
\end{proof}

\section{Lemma 1}
\begin{proof} Following the notation established in the claim of Section 2, first note that 
\begin{align*}
&\big| \text{pr}\{\theta \in \Gamma_{1 - \alpha}(S_{n}) | \theta = \theta_0 \}  - (1 - \alpha) \big|  \\
&= \big| \text{pr}\{W( \theta, S_n) \in A_{1 - \alpha} |  \theta=\theta_0 \}  -  (1 - \alpha) \big|  \\
&\leq   \big|  \text{pr}^*\{V(\theta, S_n) \in A_{1 - \alpha} | S_n = s_{\rm obs}\} - (1 - \alpha)\big|  \\
&\qquad +  \big|  \text{pr}\{W( \theta, S_n) \in A_{1 - \alpha} | \theta=  \theta_0 \} \\
&\qquad - \text{pr}^{*}\{V(\theta, S_n) \in A_{1 - \alpha} |  S_n = s_{\rm obs} \}\big| 
\end{align*} 
and by the definition of $A_{1 - \alpha}$ in (4), $\mid \text{pr}^*\{V(\theta, S_n) \in A_{1 - \alpha} \mid  S_n = s_{\rm obs}\} - (1 - \alpha)\mid   = o(\delta')$, almost surely for a pre-selected precision number, $\delta'>0$. Therefore, by Condition \ref{cond:ACC_interval}, we have $\mid  \text{pr}\{\theta \in \Gamma_{1 - \alpha}(S_{n}) \mid  \theta = \theta_0 \} - (1 - \alpha)\mid = \delta$ where $\delta = \max\{\delta_{\veps},\delta'\}$. Furthermore, if Condition \ref{cond:ACC_interval} holds almost surely, then $\mid \text{pr}\{\theta \in \Gamma_{1 - \alpha}(S_{n}) \mid \theta= \theta_0 \} - (1 - \alpha)\mid  = o(\delta)$, almost surely. 
\end{proof}

\section{Theorem 1} 
\begin{proof} If $T=T(\btheta, S_n)$ is an approximate pivot for $S_n$ then 
\begin{equation}\label{eq:AW}
		\text{pr}\{T(\btheta, S_{n}) \in A \mid \btheta = \btheta_0\} =  \int_{t \in A}
		g(t) d t \,  \{1 + o(\delta^{''})\}, 
\end{equation}
for any Borel set $A \subset {\cal S} $. Given $\btheta$ and $t$, denote the solution of $t = T( \btheta, s)$ by $s_{t, \theta}$. The density functions $g(t)$ and $f_n(s_{t, \theta}|\btheta)$ are connected by a Jacobian matrix: 
		\begin{equation}\label{eq:pivotTransformation}
		f_n(s_{t, \theta}|\btheta) |T^{(1)}( \btheta, s_{t, \theta})|^{-1} = g(t) \{1 + o(\delta^{''})\}
		\end{equation}
where $T^{(1)}( \btheta, S_n) = (\partial / \partial S_n)  T(\btheta, S_n)$. 
		
For $\theta' \sim Q_{\veps}(\cdot \mid s_{obs})$ and corresponding summary $S_n'$, 
the joint density of $(\btheta', S_{n}')$  conditional on the observed data, is
	\begin{align*}
	(\btheta', S_{n}') | S_n=  s_{\rm obs} \propto  r_{n}(\btheta) f_n( S_n \mid\btheta) K_{\veps}(S_n - s_{\rm obs}).
	\end{align*}
Let $T' = T(\btheta', S_{n}')$. With a variable transformation from $(\btheta', S_{n}')$ to $(\btheta', T')$, the joint density of $(\btheta', T')$, conditional on the observed data, is
	\begin{align*}
	(\btheta', T') |  s_{\rm obs} & \propto  r_{n}(\btheta) \left[ f_n(s_{t,\btheta}\mid\btheta) |T^{(1)}( \btheta, s_{t, \btheta})|^{-1} \right] \times\\ 
	&\qquad\qquad K_{\veps}(s_{t, \btheta}-s_{\rm obs})\\ 
	& =   r_{n}(\btheta) \left[ g(t)  \{1 + o(\delta^{''})\} \right] K_{\veps}(s_{t, \btheta}-s_{\rm obs}),
	\end{align*}
where $s_{t, \btheta}$ is the solution of  $t = T( \btheta, S_n)$ 
and the equivalence holds by (\ref{eq:pivotTransformation}). Integrating over the parameter space yields
	\begin{align*}
	T' | S_n= s_{\rm obs}  &\propto \left[ g(t)  \{1 + o(\delta^{''})\} \right] \int_{\cal{P}}  r_{n}(\theta)  K_{\veps}(s_{t, \theta}-s_{\rm obs}) d \btheta \\
	&\propto g(t) \{1 + o(\delta^{''})\},
	\end{align*} 
provided (\ref{eq:req}) holds. 
		
Now, consider $W( \btheta,S_n) = T( \theta, S_n)$ as a function of the random sample given some fixed, unknown value of $\btheta$, by (\ref{eq:AW})
	\[ \text{pr}\{W(\btheta, S_{n}) \in A \mid \btheta=  \btheta_0\} =  \int_{t \in A} g(t) d t   \{1 + o(\delta)\} .\]
If we consider $V( \btheta,S_n) = T( \theta, S_n)$ and the joint density of $(\theta', S_n')$ pairs 
then
	\[ \text{pr}^*\{V(\btheta, S_n) \in A \mid  S_n= s_{\rm obs} \} = \int_{t \in A} g(t)  dt \{1 + o(\delta^{''})\}  \]
thus satisfying Condition \ref{cond:ACC_interval}. Furthermore, by Lemma \ref{main1}, $\Gamma_{1-\alpha}(s_{\rm obs})$ in equation (\ref{eq:AB}) is a $(1-\alpha)100\%$ confidence region for $\btheta$. 
\end{proof}

\section{Corollary 1} 
\begin{proof} By Theorem \ref{thm:pivot}, it suffices to show that equation \ref{eq:apivot} is free of $t$ in each case.

\noindent {\it (a)} Suppose 
$S_n \sim g_1(S_n - \mu)$. Then $T_1 = T_1(\mu, S_n) = S_n - \mu \sim g_1(t)$ is a pivot for $S_n$. For any $(t, \mu)$ pair $s_{t,\mu}=t + \mu$.  With a change of variables $u = t + \mu - s_{obs}$ and with $r_n(\mu)\propto 1$ we have
	\begin{align*}
	\int_{\cal{P}} r_n(\mu)K_{\veps}(s_{t, \mu} - s_{\rm obs}) d\mu =  \int_{-\infty}^{\infty} K_{\veps}(u) du,
	\end{align*}
which is free of $t$.
	
\noindent {\it (b)} Suppose $S_n  \sim (1/\sigma) g_2(S_n / \sigma)$. Then $T = T(\sigma, S_n ) = S_n  / \sigma \sim g_2(t)$ is a pivot. For any $(t, \sigma)$ pair $s_{t,\sigma}= t\sigma$. With $r_n(\sigma) \propto 1/\sigma$ and with a change of variables $u = t\sigma - s_{obs}$, we have 
	\begin{align*}
	\int_{\cal{P}} r_n(\sigma) K_{\veps}(s_{t,\sigma} - s_{\rm obs}) d\sigma &=
	\int_{0}^{\infty} \frac{1}{\sigma} K_{\veps}(t\sigma - s_{\rm obs}) d\sigma \\
	&= \int_{0}^{\infty} \frac{1}{(u + s_{obs})/t}K_{\veps}(u) \frac{1}{t} du 
	\end{align*}
	which is free of $t$.
	
\noindent {\it (c)} Since we have already proven parts (a) and (b), part (c) follows provided we select $r_{n}(\theta) \propto 1/\sigma$ for $\theta = (\mu, \sigma)$. 
	
Finally, to prove the last statement of the corollary first note that the function $ H_1(S_n, x) = \int_{-\infty}^{x}g_1(S_n-u)du $ is a CD for $\mu$ when $S_n \sim g_1(S_n - \mu)$ because, for a given $S$, $H_1(S, x) $ is a distribution function on the parameter space $(-\infty, \infty)$ and given $x = \mu_0$, $H_1(S, x) \sim U(0,1)$. Similarly, the function $H_2(S^2,x) = 1 - \int_{0}^{x} g_2(S/u)du$ is a CD for $\sigma^2$ when $S_n \sim (1/\sigma)g_2(S_n/\sigma)$.  
\end{proof}

\section{Remark on degeneracy of acceptance rate}\label{sec:degen_of_AR}
A natural question is whether Theorem \ref{thm:ACC_limit_small_bandwidth} holds for a larger $\veps_n$. We claim that the answer is negative, using the following basic normal mean model as a counterexample.

Consider a univariate Gaussian model with mean $\theta$ and unit variance, and observations that are IID from the model with $\theta=\theta_0$. Let $r_{n}(\theta)$ be a normal density with mean $\mu_n$ and variance $b_n^{-2}$, where $\mu_n$ and $b_n$ are constant sequences satisfying $b_n(\mu_n-\theta_0)=O(1)$ and $b_n=o(\sqrt{n})$ as $n\rightarrow\infty$, and let $S_n$ be the sample mean. One can verify that $r_{n}$ and $S_n$ satisfy the conditions of Theorem \ref{thm:ACC_limit_small_bandwidth}. The Gaussian kernel with variance $\veps_n^2$ is used for the acceptance/rejection. Then the density of a linear transformation of $\theta \sim Q_{\veps}(\theta \mid s_{\rm obs})$ is Gaussian with a closed form  
\begin{align*}
\sqrt{n}(\theta-\hat{\theta}_{S}) \mid s_{obs} & \sim N(0,n\sigma_{\veps}^{2})
\end{align*}
where $\sigma_{\veps}^{2}=\frac{b_{n}^{-2}\Delta_{n}}{1+\Delta_{n}}$ and $\Delta_{n}=b_{n}^{2}(n^{-1}+\veps^{2})$. Also,
\begin{align*}
&\sqrt{n}(\hat{\theta}_{S}-\theta) \mid \theta_0  = \qquad \qquad \qquad \qquad \qquad \qquad \\
&\qquad \qquad \frac{1}{1+\Delta_{n}}\sqrt{n}(s_{obs}-\theta)+\frac{\sqrt{n}b_{n}^{-1}\Delta_{n}}{1+\Delta_{n}}b_{n}(\mu_{n}-\theta).
\end{align*}
By algebra, the expectation of $\sqrt{n}(\hat{\theta}_{S}-\theta) \mid \theta_0$ is $o(1)$ only when $\veps_n=o(b_{n}^{-1/2}n^{-1/4})$, and the variance is $n\sigma_{\veps}^2+o(1)$ only when $\veps_n=o(n^{-1/2})$ or $\veps_n^{-1}=o(b_n^2n^{-1/2})$. Since $b_n=o(\sqrt{n})$, both $\veps_n=o(b_{n}^{-1/2}n^{-1/4})$ and $\veps_n^{-1}=o(b_n^2n^{-1/2})$ can not hold simultaneously. Therefore Condition \ref{cond:ACC_interval} is satisfied only if $\veps_n=o(n^{-1/2})$.
	
\section{Theorems 2 and 3} \label{Appendix:Thm2and3}
The proof for Theorem 2 requires establishing Lemmas 1--5 which are given in the Supplementary Material. 
Theorem 3 is proved after establishing Lemmas 6--7 given in the Supplementary Material. 
Below, we present the additional conditions from \cite{Li2017}, necessary for these lemmas. The proofs of these technical lemmas are contained in the Supplementary Material. It is helpful to have a copy of both \cite{Li2016} and \cite{Li2017} (and their supplementary material) on hand as these proofs are rely on results from these two publications.

\subsection{Notation}\label{Appendix_notation}
For the sequence $a_n$ in Condition \ref{sum_conv}, let $a_{n,\veps}=a_{n}$ if $\lim_{n\rightarrow\infty}a_{n}\veps_{n}<\infty$ and $a_{n,\veps}=\veps_{n}^{-1}$ otherwise. Additionally, let  $\ensuremath{c_{\veps}=\lim_{n\rightarrow\infty}a_{n}\veps_{n}}$. Both $\{a_n\}$ and $c_{\veps}$ characterize how $\veps_{n}$ decreases relative to the convergence rate, $a_{n}$, of $S_{n}$ . 

Let $\ftil_{n}(\bs\mid\btheta)=N\{\bs;\bs(\btheta),A(\btheta)/a_{n}^{2}\}$ be the asymptotic distribution of the summary statistic from Condition \ref{sum_conv}. Define the standardized random variables $W_{n}(S_{n})=a_{n}A(\theta)^{-1/2}\{S_{n}-s(\theta)\}$ and $W_{\rm obs}=a_{n}A(\theta)^{-1/2}\{s_{\rm obs}-s(\theta)\}$. Finally, let $f_{W_{n}}(w\mid\theta)$ and $\ftil_{W_{n}}(w\mid\theta)$ be the density for $W_{n}(S_{n})$ when $S_{n}\sim f_{n}(\cdot\mid\theta)$ and $\ftil_n(\cdot\mid\theta),$ respectively. 


\subsection{Conditions}
\begin{condition} \label{kernel_prop}
The kernel in Algorithm \ref{alg:rejACC} satisfies 

\noindent (i) $\int vK_{\veps}(v)dv=0$; 

\noindent (ii) $\prod_{k=1}^{l}v_{i_{k}}K_{\veps}(v)dv<\infty$
		for any coordinates $(v_{i_{1}},\dots,v_{i_{l}})$ of $v$ and $l\leq p+6$;
		
\noindent (iii) $K_{\veps}(v)\propto K_{\veps}(\|v\|_{\Lambda}^{2})$ where $\|v\|_{\Lambda}^{2}=v^{T}\Lambda v$
		and $\Lambda$ is a positive-definite matrix, and $K(v)$ is a decreasing
		function of $\|v\|_{\Lambda}$; 
		
		\noindent (iv) $K_{\veps}(v)=O(\exp\{-c_{1}\|v\|^{\alpha_{1}}\})$
		for some $\alpha_{1}>0$ and $c_{1}>0$ as $\|v\|\rightarrow\infty$. 
\end{condition}

	
	\begin{condition} \label{sum_approx}
		{\it There exists $\alpha_{n}$ satisfying $\alpha_{n}/a_{n}^{2/5}\rightarrow\infty$
		and a density $r_{max}(w)$ satisfying Condition \ref{kernel_prop}, where $K_{\veps}(v)$
		is replaced with $r_{max}(w)$, such that 
		$$\sup_{\theta\in {\cal P}_{\delta} }\alpha_{n}\mid f_{W_{n}}(w\mid\theta)-\ftil_{W_{n}}(w\mid\theta)\mid\leq c_{3}r_{max}(w)$$
		for some positive constant $c_{3}$. }
	\end{condition}
	
	\begin{condition} \label{sum_approx_tail}
		{\it  For some positive constants $c_{2}$
		and $\alpha_{2}$,
		

$\sup_{\theta\in {\cal P}_{\delta} ^{C}}\ftil_{W_{n}}(w\mid\theta)=O(e^{-c_{2}\|w\|^{\alpha_{2}}})$
		as $\|w\|\rightarrow\infty$. }
	\end{condition}
	
	\begin{condition} \label{cond:likelihood_moments}
		{\it The first two moments, $\int_{{\cal S} }s\ftil_{n}(s\mid\theta)ds$
		and $\int_{\mathbb{R}^d}s^{T}s\ftil_{n}(s\mid\theta)ds$, exist. }
	\end{condition}

\bibliographystyle{nessart-number}
\bibliography{0Main-Feb2022}	
\end{document}


\maketitle


\section{Notation}
Let $N(\bx;\bmu,\Sigma)$ be the normal density at $\bx$
with mean $\bmu$ and variance $\Sigma$, and $\ftil_{n}(\bs\mid\btheta)=N\{\bs;\bs(\btheta),A(\btheta)/a_{n}^{2}\}$, the asymptotic distribution of the summary statistic.
We define $a_{n,\veps}=a_{n}$ if $\lim_{n\rightarrow\infty}a_{n}\veps_{n}<\infty$
and $a_{n,\veps}=\veps_{n}^{-1}$ otherwise, and $\ensuremath{c_{\veps}=\lim_{n\rightarrow\infty}a_{n}\veps_{n}}$,
both of which summarize how $\veps_{n}$ decreases relative to the
converging rate, $a_{n}$, of $S_{n}$ in Condition 2 below. Define
the standardized random variables $W_{n}(S_{n})=a_{n}A(\theta)^{-1/2}\{S_{n}-s(\theta)\}$
and $W_{\rm obs}=a_{n}A(\theta)^{-1/2}\{s_{\rm obs}-s(\theta_0)\}$ and $\beta_{0}=I(\theta_0)^{-1}Ds(\theta_0)^{T}A(\theta_0)^{-1}$ according to Condition 2 below. 

Let $f_{W_{n}}(w\mid\theta)$ and
$\ftil_{W_{n}}(w\mid\theta)$ be the density for $W_{n}(S_{n})$ when
$S_{n}\sim f_{n}(\cdot\mid\theta)$ and $\ftil_n(\cdot\mid\theta)$ respectively.
Let $B_{\delta}=\{\theta\mid\|\theta-\theta_{0}\|\leq\delta\}$ for
$\delta>0$. Define the initial density truncated in $B_{\delta}$,
i.e. $r_{n}(\theta)\mathbb{I}_{\theta\in B_{\delta}}/\int_{B_{\delta}}r_{n}(\theta)\,d\theta$,
by $r_{\delta}(\theta)$. Let $t(\theta)=a_{n,\veps}(\theta-\theta_{0})$
and $v(s)=\veps_{n}^{-1}(s - s_{\rm obs})$. For any $A\in\mathscr{B}^{p}$
where $\mathscr{B}^{p}$ is the Borel sigma-field on $\mathbb{R}^{p}$,
let $t(A)$ be the set $\{\phi:\phi=t(\theta)\text{ for some }\theta\in A\}$.
For a non-negative function $h(x)$, integrable in $\mathbb{R}^{l}$,
denote the normalized function $h(x)/\int_{\mathbb{R}^{l}}h(x)\,dx$
by $h(x)^{({\rm norm})}$. For a function $h(x)$, denote its gradient
by $D_{x}h(x)$, and for simplicity, omit $\theta$ from $D_{\theta}$. For a sequence $x_n$, we use the notation $x_n = \Theta(a_n)$ to mean that there exist some constants $m$ and $M$ such that $0<m<\mid x_n/a_n \mid<M<\infty$.

\section{Conditions}

{\it Condition 2.} There exists a sequence $a_{n}$, satisfying $a_{n}\rightarrow\infty$
	as $n\rightarrow\infty$, a $d$-dimensional vector $s(\btheta)$
	and a $d\times d$ matrix $A(\btheta)$, such that for $S_{n}\sim f_{n}(\cdot\mid\theta)$
	and all $\btheta\in\mathcal{P}_{0}$, 
	\[
	a_{n}\{\bS_{n}-s(\btheta)\}\rightarrow N\{0,A(\btheta)\},\mbox{ as \ensuremath{n\rightarrow\infty}},
	\]
	in distribution. We also assume that $\bs_{\rm obs}\rightarrow s(\btheta_{0})$
	in probability. Furthermore, assume
	
	 (i) $s(\btheta)$
	and $A(\btheta)\in C^{1}(\mathcal{P}_{0})$, and $A(\btheta)$ is
	positive definite for any $\btheta$; 
	
	 (ii) for any $\delta>0$ there
	exists a $\delta'>0$ such that $\|s(\btheta)-s(\btheta_{0})\|>\delta'$
	for all $\btheta$ 
	
	satisfying $\|\btheta-\btheta_{0}\|>\delta$; and
	
	 (iii) $I(\theta)\stackrel{\hbox{\tiny def}} = \left\{ \frac{\partial}{\partial\theta}s(\theta)\right\} ^{T}A(\theta)^{-1}\left\{ \frac{\partial}{\partial\theta}s(\theta)\right\} $
	has full rank at $\btheta=\btheta_{0}$.

\noindent {\it Condition 3.} There exists some $\delta_0 > 0$ such that $\mathcal{P}_{\delta_0} = \{\theta: \| \theta-\theta_0\|  < \delta_0  \} \subset \mathcal{P},$ 
For all $\theta \in \mathcal{P}_{0},$ $r_{n}(\theta) \in C^2(\mathcal{P}_{0})$ and $r_{n}(\theta_0)>0$.

\noindent {\it Condition 4.} There exists a sequence $\{\tau_{n}\}$ 
such that $\tau_{n}=o(a_n)$ and $\sup_{\btheta\in {\cal P}_{0}}\tau_{n}^{-p}r_{n}(\btheta)=O_{p}(1)$.

\noindent {\it Condition 5.} There exists constants $m$, $M$ such that $0 < m <\mid \tau_{n}^{-p}r_{n}(\btheta_{0})\mid < M < \infty$.

\noindent {\it Condition 6.} It holds that $\sup_{\btheta\in\mathbb{R}^{p}}\tau_{n}^{-1} D\{\tau_{n}^{-p}r_{n}(\btheta)\}=O_{p}(1)$.

\begin{condition} \label{kernel_prop}
	The kernel satisfies 
	
	 (i) $\int vK_{\veps}(v)dv=0$; 
	
	 (ii)$\prod_{k=1}^{l}v_{i_{k}}K_{\veps}(v)dv<\infty$
	for any coordinates $(v_{i_{1}},\dots,v_{i_{l}})$ of $v$ and $l\leq p+6$;
	
	 (iii)$K_{\veps}(v)\propto K_{\veps}(\|v\|_{\Lambda}^{2})$ where $\|v\|_{\Lambda}^{2}=v^{T}\Lambda v$
	and $\Lambda$ is a positive-definite matrix, and $K(v)$ is a decreasing
	function of $\|v\|_{\Lambda}$; (iv) $K_{\veps}(v)=O(\exp\{-c_{1}\|v\|^{\alpha_{1}}\})$
	for some $\alpha_{1}>0$ and $c_{1}>0$ as $\|v\|\rightarrow\infty$. 
\end{condition}

\begin{condition} \label{sum_approx}
	There exists $\alpha_{n}$ satisfying $\alpha_{n}/a_{n}^{2/5}\rightarrow\infty$
	and a density $r_{max}(w)$ satisfying Condition \ref{kernel_prop}(ii)--(iii) where $K_{\veps}(v)$
	is replaced with $r_{max}(w)$, such that $\sup_{\theta\in B_{\delta}}\alpha_{n}\mid f_{W_{n}}(w\mid\theta)-\ftil_{W_{n}}(w\mid\theta)\mid\leq c_{3}r_{max}(w)$
	for some positive constant $c_{3}$. 
\end{condition}

\begin{condition} \label{sum_approx_tail}
	The following statements hold: 
	
	(i) $r_{max}(w)$ satisfies
	Condition \ref{kernel_prop}(iv); and 
	
	(ii) $\sup_{\theta\in B_{\delta}^{C}}\ftil_{W_{n}}(w\mid\theta)=O(e^{-c_{2}\|w\|^{\alpha_{2}}})$
	as $\|w\|\rightarrow\infty$ for some positive constants $c_{2}$
	and $\alpha_{2}$, and $A(\theta)$ is bounded in ${\cal P}$. 
\end{condition}

\begin{condition} \label{cond:likelihood_moments}
	The first two moments, $\int_{\mathbb{R}^{d}}s\ftil_{n}(s\mid\theta)ds$
	and $\int_{\mathbb{R}^d}s^{T}s\ftil_{n}(s\mid\theta)ds$, exist. 
\end{condition}

\section{Proof for Theorem 2}

\noindent Let $\tilde{Q}(\theta\in A\mid s)=\int_{A}r_{\delta}(\theta)\ftil_{n}(s\mid\theta)\,d\theta/\int_{\mathbb{R}^{p}}r_{\delta}(\theta)\ftil_{n}(s\mid\theta)\,d\theta$. 

\begin{lemma}\label{Alemma1} Assume Condition \ref{par_true}--\ref{sum_approx}. If $\veps_{n}=O(a_{n}^{-1})$, for any fixed $\nu\in\mathbb{R}^{d}$
	and small enough $\delta$, 
	\[
	\sup_{A\in\mathfrak{B}^{p}}\left|\tilde{Q}\{a_{n}(\theta-\theta_{0})\in A\mid s_{\rm obs}+\veps_{n}\nu\}-\int_{A}N[t;\beta_{0}\{A(\theta_{0})^{1/2}W_{\rm obs}+c_{\veps}\nu\},I(\theta_{0})^{-1}]dt\right|\rightarrow0,
	\]
	in probability as $n\rightarrow\infty$, where $\beta_{0}=I(\theta_{0})^{-1}Ds(\theta_{0})^{T}A(\theta_{0})^{-1}$.
\end{lemma}

\noindent {\it Proof of Lemma \ref{Alemma1}:} 
	This result generalizes Lemma A1 in ~\cite{Li2017}. With Lemma A1 from ~\cite{Li2017}, it is sufficient to show that 
	\[
	\sup_{A\in\mathfrak{B}^{p}}\mid\tilde{Q}\{t(\theta)\in A\mid s_{\rm obs}+\veps_{n}\nu\}-\tilde{\Pi}\{t(\theta)\in A\mid s_{\rm obs}+\veps_{n}\nu\}\mid=o_{P}(1),
	\]
	where $\tilde{\Pi}$ denotes the posterior distribution with prior $\pi_{\delta}(\theta)$ and likelihood $\tilde{f}_n(s \mid \theta)$. 
With the transformation $t=t(\theta)$
	and $v=v(s)$, the left hand side of the above equation can be written
	as 
	\begin{eqnarray}
	&&\sup_{A\in\mathfrak{B}^{p}}\mid\frac{\int_{A}r_{\delta}(\theta_0+a_{n}^{-1}t)\ftil_{n}(s_{{\rm obs}}+\veps_{n}\nu\mid\theta_0+a_{n}^{-1}t)dt}{\int_{\mathbb{R}^{p}}r_{\delta}(\theta_0+a_{n}^{-1}t)\ftil_{n}(s_{{\rm obs}}+\veps_{n}\nu\mid\theta_0+a_{n}^{-1}t)dt}-\hfill  \label{eq1} \\
	&&\hspace{3cm}\frac{\int_{A}\pi_{\delta}(\theta_0+a_{n}^{-1}t)\ftil_{n}(s_{{\rm obs}}+\veps_{n}\nu\mid\theta_0+a_{n}^{-1}t)dt}{\int_{\mathbb{R}^{p}}\pi_{\delta}(\theta_0+a_{n}^{-1}t)\ftil_{n}(s_{{\rm obs}}+\veps_{n}\nu\mid\theta_0+a_{n}^{-1}t)dt}\mid. \notag 
	\end{eqnarray}
	For a function $\tau:\mathbb{R}^{p}\rightarrow\mathbb{R},$ define
	the following auxiliary functions,
	\begin{eqnarray*}
		\phi_{1}\{\tau(\theta);n\} & = & \frac{\int_{t(B_{\delta})}|\tau(\theta_0+a_{n}^{-1}t)-\tau(\theta)|\ftil_{n}(s_{{\rm obs}}+\veps_{n}\nu\mid\theta_0+a_{n}^{-1}t)\,dt}{\int_{t(B_{\delta})}\tau(\theta_0+a_{n}^{-1}t)\ftil_{n}(s_{{\rm obs}}+\veps_{n}\nu\mid\theta_0+a_{n}^{-1}t)\,dt},\\
		\phi_{2}\{\tau(\theta);n\} & = & \frac{\tau(\theta)\int_{t(B_{\delta})}\ftil_{n}(s_{{\rm obs}}+\veps_{n}\nu\mid\theta_0+a_{n}^{-1}t)dt}{\int_{t(B_{\delta})}\tau(\theta_0+a_{n}^{-1}t)\ftil_{n}(s_{{\rm obs}}+\veps_{n}\nu\mid\theta_{0}+a_{n}^{-1}t)dt}.
	\end{eqnarray*}
	Then by adding and subtracting $\phi_{2}\{\tau_{n}^{-p}r_{\delta}(\theta);n\}\phi_{2}\{\pi(\theta);n\}$
	in the absolute sign of \eqref{eq1}, \eqref{eq1} can be bounded
	by 
	\begin{eqnarray*}
		\phi_{1}\{\tau_{n}^{-p}r_{\delta}(\theta);n\}+\phi_{1}\{\pi(\theta);n\}\phi_{2}\{\tau_{n}^{-p}r_{\delta}(\theta);n\}+\phi_{1}\{\tau_{n}^{-p}r_{\delta}(\theta);n\}\phi_{2}\{\pi(\theta);n\}+\phi_{1}\{\pi(\theta);n\}.
	\end{eqnarray*}
	Consider a class of function $\tau(\theta)$ satisfying the following
	conditions: 
	
	\noindent (1) There exists a series $\{k_{n}\}$, such that $\sup_{\theta\in\mathcal{P}_{0}}\|k_{n}^{-1}D\tau(\theta)\|<\infty$
	and $k_{n}=o(a_{n});$ 
	
    \noindent (2)	$\tau(\theta_{0})>0$ and $\tau(\theta)\in C^{1}(B_{\delta}).$

	By Conditions \ref{par_true}--\ref{initial_gradient}, $\tau_{n}^{-p}r_{\delta}(\theta)$
	and $\pi_{\delta}(\theta)$ belong to the above class. Then if $\phi_{1}\{\tau(\theta);n\}$
	is $o_{p}(1)$ and $\phi_{2}\{\tau(\theta);n\}$ is $O_{p}(1)$, \eqref{eq1}
	is $o_{p}(1)$ and the lemma holds. 
	
	First, from the properties of $\tau(\theta)$ mentioned above, there exists an open set $\omega\subset B_{\delta}$
	such that $\inf_{\theta\in\omega}\tau(\theta)>c_{1}$, for a constant
	$c_{1}>0$. Then for $\phi_{2}\{\tau(\theta);n\}$, it is bounded
	by 
	\[
	\frac{\tau(\theta)}{c_{1}\int_{t(\omega)}\ftil_{n}(s_{{\rm obs}}+\veps_{n}\nu\mid\theta_{0}+a_{n}^{-1}t)^{(norm)}dt},
	\]
	where $h(x)^{(norm)}$ represents the normalized version of $h(x)$.
	From equation (7) in the supplementary material of \cite{Li2016},
	$\ftil_{n}(s_{{\rm obs}}+\veps_{n}\nu\mid\theta_0+a_{n}^{-1}t)$
	can be written in the following form, 
	\begin{eqnarray}
	a_{n}^{d}\ftil_{n}(s_{{\rm obs}}+\veps_{n}\nu\mid\theta_0+a_{n}^{-1}t)=\frac{1}{\|B_{n}(t)\|^{1/2}}N[C_{n}(t)\{A_{n}(t)t-b_{n}\nu-c_{2}\};\theta_0,I_{d}],\label{eq2}
	\end{eqnarray}
	where $A_{n}(t)$ is a series of $d\times p$ matrix functions, $\{B_{n}(t)\}$
	and $\{C_{n}(t)\}$ are a series of $d\times d$ matrix functions,
	$b_{n}$ converges to a non-negative constant and $c_{2}$ is a constant,
	and the minimum of absolute eigenvalues of $A_{n}(t)$ and eigenvalues
	of $B_{n}(t)$ and $C_{n}(t)$ are all bounded and away from $0$.
	Then for fixed $\nu$, by continuous mapping, \eqref{eq2} is away
	from zero with probability one. Therefore $\phi_{2}\{\tau(\theta);n\}=O_{P}(1)$.
	
	Second, by Taylor expansion, $\tau(\theta_0+a_{n}^{-1}t)=\tau(\theta_0)+a_{n}^{-1}D\tau(\theta_0+e_{t}t)t$,
	where $\|e_{t}\|\leq a_{n}^{-1}$. Then $\phi_{1}\{\tau(\theta);n\}$ is equal to
	\begin{eqnarray}
    &\frac{k_{n}\phi_{2}\{\tau(\theta);n\}}{a_{n}\tau(\theta)}\frac{\int_{t(B_{\delta})}|k_{n}^{-1}D\tau(\theta_0+e_{t}t)t|\ftil_{n}(s_{{\rm obs}}+\veps_{n}\nu\mid\theta_0+a_{n}^{-1}t)\,dt}{\int_{t(B_{\delta})}\ftil_{n}(s_{{\rm obs}}+\veps_{n}\nu\mid\theta_0+a_{n}^{-1}t)\,dt}\nonumber \\
	& \leq  \frac{k_{n}\phi_{2}\{\tau(\theta);n\}}{a_{n}\tau(\theta)}\sup_{\theta\in B_{\delta}}\|k_{n}^{-1}D\tau(\theta)\|\frac{\int_{t(B_{\delta})}\|t\|a_{n}^{d}\ftil_{n}(s_{{\rm obs}}+\veps_{n}\nu\mid\theta_0+a_{n}^{-1}t)dt}{\int_{t(B_{\delta})}a_{n}^{d}\ftil_{n}(s_{{\rm obs}}+\veps_{n}\nu\mid\theta_0+a_{n}^{-1}t)\,dt},\label{eq3}
	\end{eqnarray}
	where the inequality holds by the triangle inequality. By the expression
	\eqref{eq2} and Lemma 7 in the supplementary material of ~\cite{Li2016},
	the right hand side of \eqref{eq3} is $O_{P}(1)$. Then together
	with $\phi_{2}\{\tau(\theta);n\}=\Theta_{P}(1)$, $\phi_{1}\{\tau(\theta);n\}=o_{P}(1).$
	Therefore the Lemma holds.
	\hfill{$\square$} 

\pagebreak 

Define the joint density of $(\theta,s)$ in Algorithm \ref{alg:rejACC} and its approximation, where the s-likelihood is replaced by its Gaussian
limit and $r_{n}(\theta)$ by its truncation, by $q_{\veps}(\theta,s)$
and $\tilde{q}_{\veps}(\theta,s)$. Then
\begin{align*}
q_{\veps}(\theta,s) & =\frac{r_{n}(\theta)f_{n}(s|\theta)K_{\veps_{n}}(s-s_{\rm obs})}{\int_{\mathbb{R}^{p}\times\mathbb{R}^{d}}r_{n}(\theta)f_{n}(s|\theta)K_{\veps_{n}}(s-s_{\rm obs})\,d\theta ds},\\
\tilde{q}_{\veps}(\theta,s) & =\frac{r_{\delta}(\theta)\ftil_{n}(s|\theta)K_{\veps_{n}}(s-s_{\rm obs})}{\int_{\mathbb{R}^{p}\times\mathbb{R}^{d}}r_{\delta}(\theta)\ftil_{n}(s|\theta)K_{\veps_{n}}(s-s_{\rm obs})\,d\theta ds}.
\end{align*}
Let $\tilde{Q}_{\veps}(\theta\in A\mid s_{\rm obs})$ be the approximate confidence distribution function equal to $\int_{A}\int_{\mathbb{R}^{d}}\tilde{q}_{\veps}(\theta,s)\,dsd\theta$. 
With the transformation $t=t(\theta)$ and $v=v(s)$, let $$\tilde{q}_{\veps,t\nu}(t,v)=\tau_{n}^{-p}r_{\delta}(\theta_0+a_{n,\veps}^{-1}t)\ftil_{n}(s_{\rm obs}+\veps_{n}\nu\mid\theta_0+a_{n,\veps}^{-1}t)K_{\veps}(\nu)$$
be the transformed and unnormalized $\tilde{q}_{\veps}(\theta,s)$, and
$$\tilde{q}_{A,tv}(h)=\int_{A}\int_{\mathbb{R}^{d}}h(t,v)\tilde{q}_{\veps,t\nu}(t,v)\,dvdt$$
for any function $h(\cdot,\cdot)$ in $\mathbb{R}^{p}\times\mathbb{R}^{d}$.
Denote the factor of $\tilde{q}_{\veps,t\nu}(t,v)$, $\tau_{n}^{-p}r_{\delta}(\theta_0+a_{n,\veps}^{-1}t)$,
by $\gamma_{n}(t)$. Let $\gamma=\lim_{n\rightarrow\infty}\tau_{n}^{-p}r_{\delta}(\theta)$
and $\gamma(t)=\lim_{n\rightarrow\infty}\tau_{n}^{-p}r_{\delta}(\theta_0+\tau_{n}^{-1}t)$,
the limits of $\gamma_{n}(t)$ when $a_{n,\veps}=a_{n}$ and
$a_{n,\veps}=\tau_{n}$ respectively. By Condition \ref{initial_upper} and \ref{initial_lower},
$\gamma(t)$ exists and $\gamma$ is non-zero with positive probability.

Next several functions of $t$ and $v$ defined in ~\cite[proofs for Section 3.1]{Li2017}
and relate to the limit of $\tilde{q}_{\veps,t\nu}(t,v)$ are used, including
$g(v;A,B,c)$, $g_{n}(t,v)$, $G_{n}(v)$ and $g_{n}'(t,v)$. 

Furthermore
several functions defined by integration as following are used: for
any $A\in\mathfrak{B}^{p}$, let $$g_{A,r}(h)=\int_{\mathbb{R}^{d}}\int_{t(A)}h(t,v)\gamma_{n}(t)g_{n}(t,v)\,dtdv,$$
$$G_{n,r}(v)=\int_{t(B_{\delta})}\gamma_{n}(t)g_{n}(t,v)\,dt,$$ 
$$q_{A}(h)=\int_{A}\int_{\mathbb{R}^{d}}h(\theta,s)r_{n}(\theta)f_{n}(s\mid\theta)K_{\veps}(s-s_{\rm obs})\veps_{n}^{-d}\,dsd\theta,$$
$$\tilde{q}_{A}(h)=\int_{A}\int_{\mathbb{R}^{d}}h(\theta,s)r_{\delta}(\theta)\ftil_{n}(s\mid\theta)K_{\veps}(s-s_{\rm obs})\veps_{n}^{-d}\,dsd\theta,$$
which generalize those defined in ~\cite[proofs for Section 3.1]{Li2017}
for the case $r_{n}(\theta)=\pi(\theta)$.

\begin{lemma}\label{Alemma2} Assume Condition \ref{par_true}--\ref{kernel_prop}. If $\veps_{n}=o(a_{n}^{-1/2})$, then 
	\begin{enumerate}
		\item[(i)] $\int_{\mathbb{R}^{d}}\int_{t(B_{\delta})}|\tilde{q}_{\veps,t\nu}(t,\nu)-\gamma_{n}(t)g_{n}(t,\nu)|\,dtd\nu=o_{p}(1)$;
		\item[(ii)] $g_{B_{\delta},r}(1)=\Theta_{P}(1);$ 
		\item[(iii)] $\tilde{q}_{B_{\delta},tv}(t^{k_{1}}v^{k_{2}})/\tilde{q}_{B_{\delta},tv}(1)=g_{B_{\delta},r}(t^{k_{1}}v^{k_{2}})/g_{B_{\delta},r}(1)+O_{P}(a_{n,\veps}^{-1})+O_{P}(a_{n}^{2}\veps_{n}^{4})$
		for pairs $(k_{1}, k_{2}) = (0,0), (1,0), (1,1), (0,1)$, and $(0,2)$; 
		\item[(iv)] $\tilde{q}_{B_{\delta}}(1)=\tau_{n}^{p}a_{n,\veps}^{d-p}\left\{ \int_{t(B_{\delta})}\int_{\mathbb{R}^{d}}\gamma_{n}(t)g_{n}(t,\nu)d\tau d\nu+O_{P}(a_{n,\veps}^{-1})+O_{P}(a_{n}^{2}\veps_{n}^{4})\right\} $. 
	\end{enumerate}\end{lemma}
\noindent {\it Proof of Lemma \ref{Alemma2}:} 
	These results generalize parts of Lemma A2 in ~\cite{Li2017} (corresponding to items $(i)$ and $(ii)$ above) and Lemma 5 in
	~\cite{Li2016} (corresponding to items $(iii)$ and $(iv)$ above). 
	
	To prove part $(i)$, note that in Lemma A2 of ~\cite{Li2017}  $\gamma_{n}(t)=\pi(\theta_0+a_{n,\veps}^{-1}t)$,
	and $(i)$ holds by expanding $\tilde{q}_{\veps,t\nu}(t,v)$ according to
	the proof of Lemma 5 of ~\cite{Li2016}. 
	For $\gamma_{n}(t)=\tau_{n}^{-p}r_{\delta}(\theta_0+a_{n,\veps}^{-1}t)$ this can be similarly proved by changing the terms involving $\pi(\theta)$ in equations (10) and (11) in the supplements of ~\cite{Li2016}. Equation (10)
	is replaced by 
	\[
	\frac{\gamma_{n}(t)}{\mid A(\theta+a_{n,\veps}^{-1}t)\mid^{1/2}}=\frac{\gamma_{n}(t)}{\mid A(\theta)\mid^{1/2}}+a_{n,\veps}^{-1}\gamma_{n}(t)D\frac{1}{\mid A(\theta+e_{t})\mid^{1/2}}t,
	\]
	where $\|e_{\tau}\|\leq\delta$, and this leads to replacing $\pi(\theta_0)\int_{\tau(B_{\delta})\times\mathbb{R}^{d}}g_{n}(t,\nu)dtd\nu$
	in equation (11) by $\int_{\tau(B_{\delta})\times\mathbb{R}^{d}}\gamma_{n}(t)g_{n}(t,\nu)dtd\nu$.
	These changes have no effect on the arguments therein since $\sup_{t\in t(B_{\delta})}\gamma_{n}(t)=O_{P}(1)$
	by Condition \ref{initial_upper}. Therefore $(i)$ holds.
	
	For (ii), By Condition \ref{initial_lower} and Lemma A2 of ~\cite{Li2017}, there exists a $\delta'<\delta$
	such that $\inf_{t\in t(B_{\delta'})}\gamma_{n}(t)=\Theta_{p}(1)$
	and $\int_{\mathbb{R}^{d}}\int_{t(B_{\delta'})}g_{n}(t,\nu)\,dtdv=\Theta_{p}(1)$.
	Then since $g_{B_{\delta},r}(1)\geq\inf_{t\in t(B_{\delta'})}\gamma_{n}(t)\int_{\mathbb{R}^{d}}\int_{t(B_{\delta'})}g_{n}(t,\nu)\,dtd\nu$,
	(ii) holds.
	
	For $(iii)$, if $(k_1,k_2)=(1,0)$ then $\tilde{q}_{B_{\delta},tv}(t)/\tilde{q}_{B_{\delta},tv}(1)$
	can be expanded by following the arguments in the proof of Lemma 5
	of ~\cite{Li2016}. For the other pairs of $(k_1,k_2)$, $\tilde{q}_{B_{\delta},tv}(t^{k_{1}}v^{k_{2}})/\tilde{q}_{B_{\delta},tv}(1)$,
	can be expanded similarly as in the proof of Lemma 4 from ~\cite{Li2017}.
	
	For $(iv)$, $\gamma_{n}(t)$ plays the same role as $\pi(\theta)$
	in the proof of Lemma 5 in ~\cite{Li2016}, and the arguments therein
	can be followed exactly. The term $\tau_{n}^{p}$ is from the definition
	of $\gamma_{n}(t)$ that $r_{n}(\theta_0+a_{n,\veps}^{-1}t)=\tau_{n}^{p}\gamma_{n}(t)$.
	\hfill{$\square$}


Recall the definition of the estimator $\theta_{\veps} = \int \theta dQ_{\veps}(\theta \mid s_{obs}) d\theta$. Define the expectation of $\theta$ with distribution $\tilde{Q}_{\veps}(\theta\in A\mid s_{\rm obs})$
as $\tilde{\theta}_{\veps}$ and the expectation of the regression adjusted values, $\theta^*$ 
with density $\tilde{q}_{\veps}(\theta,s)$ as $\tilde{\theta}_{\veps}^{*}$.
Let $E_{G,r}(\cdot)$ be the expectation with the density $G_{n}(v)^{({\rm norm})}$,
and $E_{G,r}\{h(v)\}$ can be written as $g_{B_{\delta},r}\{h(v)\}/g_{B_{\delta},r}(1)$.
Let $\psi(\nu)=k_{n}^{-1}\beta_{0}\{A(\theta_{0})^{1/2}W_{\rm obs}+a_{n}\veps_{n}\nu\}$,
where $k_{n}=1$, if $c_{\veps}<\infty$, and $a_{n}\veps_{n}$,
if $c_{\veps}=\infty$. 


\begin{lemma}\label{Alemma3} Assume Condition \ref{par_true}--\ref{initial_gradient} and \ref{kernel_prop}. Then if $\veps_{n}=o(a_{n}^{-1/2})$, 
	\begin{enumerate}
		\item[(i)] $\tilde{\theta}_{\veps}=\theta_{0}+a_{n}^{-1}\beta_{0}A(\theta_{0})^{1/2}W_{\rm obs}+\veps_{n}\beta_{0}E_{G_{n},r}(\nu)+r_{1}$,
		where $r_{1}=o_{P}(a_{n}^{-1})$; 
		\item[(ii)] $\tilde{\theta}_{\veps}^{*}=\theta_{0}+a_{n}^{-1}\beta_{0}A(\theta_{0})^{1/2}w_{\rm obs}+\veps_{n}(\beta_{0}-\beta_{\veps})E_{G_{n},r}(\nu)+r_{2}$,
		where $r_{2}=o_{P}(a_{n}^{-1})$. 
	\end{enumerate}\end{lemma}


\noindent {\it Proof of Lemma \ref{Alemma3}:} 
	These results generalize Lemma A3(c) and Lemma A5(c) in ~\cite{Li2017} in the sense of permitting use of a data-dependent $r_n(\theta)$, however here we are only considering $\veps_n = o(a_n^{-1/2})$ in contrast to Lemma A5(c) in ~\cite{Li2017} which assumes the less stringent condition that $\veps_n = o(a_n^{-3/5})$.
	
	With the transformation $t=t(\theta)$, by Lemma 2, if $\veps_{n}=o(a_{n}^{-1/2})$,
	\begin{eqnarray}
	\begin{cases}
	\tilde{\theta}_{\veps}=\theta_{0}+a_{n,\veps}^{-1}\tilde{q}_{B_{\delta},t\nu}(t)/\tilde{q}_{B_{\delta},t\nu}(1)=\theta_{0}+a_{n,\veps}^{-1}g_{B_{\delta},r}(t)/g_{B_{\delta},r}(1)+o_{p}(a_{n}^{-1}),\\
	\tilde{\theta}_{\veps}^{*}=\theta_{0}+a_{n,\veps}^{-1}\tilde{q}_{B_{\delta},t\nu}(t)/\tilde{q}_{B_{\delta},t\nu}(1)-\veps_{n}\beta_{\veps}\tilde{q}_{B_{\delta},t\nu}(\nu)/\tilde{q}_{B_{\delta},t\nu}(1)\\
	\hspace{5mm} =\theta_{0}+a_{n,\veps}^{-1}g_{B_{\delta},r}(t)/g_{B_{\delta},r}(1)-\veps_{n}\beta_{\veps}E_{a_{n},r}(\nu)+o_{p}(a_{n}^{-1}),
	\end{cases}\label{eq4}
	\end{eqnarray}
	where the remainder term comes from the fact that $(a_{n,\veps}^{-1}+\veps_{n})\left\{ O_{p}(a_{n,\veps}^{-1})+O_{p}(a_{n}^{2}\veps_{n}^{4})\right\} =o_{p}(a_{n}^{-1})$.
	
	First the leading term of $g_{B_{\delta},r}(t\nu^{k})$ is derived
	for $k=0$ or $1$. The case of $k=1$ will be used later. Let $t'=t-\psi(\nu)$,
	then 
	\begin{align*}
	g_{B_{\delta},r}(t\nu^{k_{2}}) & =\int_{\mathbb{R}^{d}}\int_{t(B_{\delta})}\{t'+\psi(\nu)\}\nu^{k_{2}}\gamma_{n}(t)g_{n}(t,\nu)\,dtd\nu\\
	& =\int_{\mathbb{R}^{d}}\psi(\nu)\nu^{k_{2}}G_{n,r}(\nu)\,d\nu+\int_{\mathbb{R}^{d}}\int_{t(B_{\delta})}t'\nu^{k_{2}}\gamma_{n}(t)g_{n}(t,\nu)\,dtd\nu.
	\end{align*}
	By matrix algebra, it is straightforward to show that $g_{n}(t,v)=N\{t;\psi(v),k_{n}^{-2}I(\theta_{0})^{-1}\}G_{n}(v)$.
	Then with the transformation $t'$, we have
	\begin{align*}
	& g_{B_{\delta},r}(t\nu^{k_{2}})-\int_{\mathbb{R}^{d}}\psi(\nu)\nu^{k_{2}}G_{n,r}(\nu)\,d\nu\\
	= & \int_{\mathbb{R}^{d}}\int_{t(B_{\delta})-\psi(\nu)}t'\nu^{k_{2}}\gamma_{n}\{\psi(\nu)+t'\}N\left\{ t';0,k_{n}^{-2}I(\theta_{0})^{-1}\right\} G_{n}(\nu)\,dt'd\nu.
	\end{align*}
	By applying the Taylor expansion on $\gamma_{n}\{\psi(\nu)+t'\}$,
	the right hand side of the above equation is equal to 
	\begin{eqnarray}
	&  & \int_{\mathbb{R}^{d}}\int_{t(B_{\delta})-\psi(\nu)}t'N\{t';0,k_{n}^{-2}I(\theta_{0})^{-1}\}\,dt'\cdot\gamma_{n}\{\psi(\nu)\}\nu^{k_{2}}G_{n}(\nu)\,d\nu\nonumber \\
	&  & +\int_{\mathbb{R}^{d}}\int_{t(B_{\delta})-\psi(\nu)}t'{}^{2}D_{t}\gamma_{n}\{\psi(\nu)+e_{t}\}N\{t';0,k_{n}^{-2}I(\theta_{0})^{-1}\}\,dt'\cdot\nu^{k_{2}}G_{n}(\nu)d\nu\nonumber \\
	& = & k_{n}^{-1}\int_{\mathbb{R}^{d}}\int_{Q_{v}}t''N\{t'';0,I(\theta_{0})^{-1}\}\,dt''\cdot\gamma_{n}\{\psi(\nu)\}\nu^{k_{2}}G_{n}(\nu)\,d\nu\nonumber \\
	&  & +k_{n}^{-2}\int_{\mathbb{R}^{d}}\int_{Q_{v}}t''^{2}D_{t}\gamma_{n}\{\psi(\nu)+e_{t}\}N\{t'';0,I(\theta_{0})^{-1}\}\,dt''\cdot\nu^{k_{2}}G_{n}(\nu)\,d\nu,\label{eq5}
	\end{eqnarray}
	where $Q_{v}=\left\{ a_{n}(\theta-\theta_{0})-k_{n}\psi(\nu)\mid\theta\in B_{\delta}\right\} $
	and $t''=k_{n}t'$. 
	Since $Q_{v}$ can be written as $\left\{ a_{n}(\theta-\theta_{0}-\beta_{0}\veps_{n}\nu)-\beta_{0}A(\theta_{0})^{1/2}W_{\rm obs}\mid\theta\in B_{\delta}\right\} $,
	it converges to $\mathbb{R}^{p}$ for any fixed $v$ with probability one using the dominated convergence theorem. 
	Then $\int_{Q_{v}}t''N\{t'';0,\tau(\theta_{0})^{-1}\}\,dt''=o_{P}(1)$
	for fixed $v$, and by the continuous mapping theorem and Condition \ref{initial_upper}, the
	first term in the right hand side of \eqref{eq5} is of the order
	$o_{p}(k_{n}^{-1})$. The second term is bounded by 
	\[
	k_{n}^{-2}\sup_{t\in\mathbb{R}}\|D_{t}\gamma_{n}(t)\|\int_{\mathbb{R}^{p}}\|t''\|^{2}N\{t'';0,I(\theta_{0}^{-1})\}\,dt''\int_{\mathbb{R}^{d}}\nu^{k_{2}}G_{n}(\nu)\,d\nu,
	\]
	which is of the order $O_{p}(k^{-2}\tau_{n}/a_{n,\veps})$ by Condition \ref{initial_gradient}.
	Therefore 
	\begin{align}
	g_{B_{\delta},r}(t\nu^{k_{2}}) & =\int_{\mathbb{R}^{d}}\psi(\nu)\nu^{k_{2}}G_{n}(\nu)d\nu+o_{P}(k_{n}^{-1}).\label{eq6}
	\end{align}
	By algebra, $k_{n}=a_{n,\veps}^{-1}a_{n}$, and 
	\begin{eqnarray}
	&  & \int_{\mathbb{R}^{d}}\psi(\nu)\nu^{k_{2}}G_{n}(\nu)d\nu\nonumber \\
	& = & a_{n,\veps}\beta_{0}\{a_{n}^{-1}A(\theta_{0})^{1/2}W_{\rm obs}\int_{\mathbb{R}^{d}}\nu^{k_{2}}G_{n,r}(\nu)\,d\nu+\veps_{n}\int_{\mathbb{R}^{d}}\nu^{k_{2}+1}G_{n,r}(\nu)\,d\nu\}.\label{eq7}
	\end{eqnarray}
	Then ($i$) and ($ii$) in the Lemma holds by plugging the expansion
	of $g_{B_{\delta},r}(t)$ into \eqref{eq4}.
	
	\hfill{$\square$} 
	

\begin{lemma}\label{Alemma3.5} Assume Condition \ref{par_true}, \ref{initial_upper}, \ref{sum_conv}--\ref{sum_approx_tail}. Then as $n\rightarrow\infty$, 
	\begin{enumerate}
		\item[(i)] For any $\delta<\delta_{0}$, $r_{B_{\delta}^{c}}(1)$ and $\tilde{q}_{B_{\delta}^{c}}(1)$
		are $o_{p}(\tau_{n}^{p})$. More specifically, they are of the order
		$O_{p}\left(\tau_{n}^{p}e^{-a_{n,\veps}^{\alpha_{\delta}}c_{\delta}}\right)$
		for some positive constants $c_{\delta}$ and $\alpha_{\delta}$ depending
		on $\delta$.
		\item[(ii)] $q_{B_{\delta}}(1)=\tilde{q}_{B_{\delta}}(1)\{1+O_{p}(\alpha_{n}^{-1})\}$
		and $\sup_{A\subset B_{\delta}}|q_{A}(1)-\tilde{q}_{A}(1)|/\tilde{q}_{B_{\delta}}(1)=O_{p}(\alpha_{n}^{-1})$; 
		\item[(iii)] if $\veps_{n}=o(a_{n}^{-1/2})$, then $\tilde{q}_{B_{\delta}}(1)$ and
		$r_{B_{\delta}}(1)$ are $\Theta_{P}(\tau_{n}^{p}a_{n,\veps}^{d-p})$,
		and thus $\tilde{q}_{\mathcal{P}_{0}}(1)$ and $q_{\mathcal{P}_{0}}(1)$
		are $\Theta_{P}(\tau_{n}^{p}a_{n,\veps}^{d-p})$; 
		\item[(iv)] if $\veps_{n}=o(a_{n}^{-1/2})$, $\theta_{\veps}=\tilde{\theta}_{\veps}+o_{p}(a_{n,\veps}^{-1}).$
		If $\veps_{n}=o(a_{n}^{-3/5}),$ $\theta_{\veps}=\tilde{\theta}_{\veps}+o_{P}(a_{n}^{-1}).$
	\end{enumerate} \end{lemma}
\noindent {\it Proof of Lemma \ref{Alemma3.5}:} 
	This generalizes Lemma 2 in the supplements of ~\cite{Li2017}. The arguments therein
	can be followed exactly, by Condition \ref{initial_upper} and the fact that regarding $\pi(\theta)$,
	only the condition $\sup_{\theta\in\mathbb{R}^{p}}\pi(\theta)<\infty$
	is used.
	
	\hfill{$\square$} 
	
\begin{lemma}\label{Alemma4} 
Assume Condition \ref{par_true}, \ref{initial_upper}, \ref{sum_conv}--\ref{sum_approx_tail}. 
	\begin{enumerate}
		\item[(i)] For any $\delta<\delta_{0}$, $Q_{\veps}(\theta\in B_{\delta}^{c}\mid s_{\rm obs})$
		and $\tilde{Q}_{\veps}(\theta\in B_{\delta}^{c}\mid s_{\rm obs})$ are $o_{p}(1)$; 
		\item[(ii)] There exists some $\delta<\delta_{0}$ such that 
		\[
		\sup_{A\in\mathfrak{B}^{p}}|Q_{\veps}(\theta\in A\cap B_{\delta}\mid s_{\rm obs})-\tilde{Q}_{\veps}(\theta\in A\cap B_{\delta}\mid s_{\rm obs})|=o_{p}(1);
		\]
		\item[(iii)] $a_{n,\veps}(\theta_{\veps}-\tilde{\theta}_{\veps})=o_{p}(1)$ . 
	\end{enumerate}\end{lemma}
\noindent {\it Proof of Lemma \ref{Alemma4}:} 
	This lemma generalizes Lemma A3 of~\cite{Li2017}. The proof of Lemma A3 of~\cite{Li2017} only needs Lemma 3 and 5 from~\cite{Li2016} to hold. The result that  $q_{B_{\delta}^{c}}\{h(\theta)\}=O_{p}(\tau_{n}^{p}e^{-a_{n,\veps}^{\alpha_{\delta}}c_{\delta}})$
	for some positive constants $\alpha_{\delta}$ and $c_{\delta}$, which generalizes the case of $r_{n}(\theta)=\pi(\theta)$ in Lemma 3 of~\cite{Li2016}, holds by Condition \ref{initial_upper}, since the latter only uses the fact that 
	$\sup_{\theta\in B_{\delta}^{c}}\pi(\theta)<\infty$.
	Then the arguments in the proof of Lemma 3 in~\cite{Li2016} can be followed exactly, despite the term $\tau_{n}^{p}$ that is not included in the order of $\pi_{B_{\delta}^{c}}\{h(\theta)\}$, since $Q_{\veps}(\theta\in A\mid s_{{\rm obs}})$
	is the ratio $q_{A}(1)/q_{\mathbb{R}^{p}}(1)$. Since Lemma 5 in~\cite{Li2016} has been generalized by Lemma \eqref{Alemma2} above, the arguments of the proof of Lemma A3 in~\cite{Li2017} can be followed exactly.
	\hfill{$\square$} \\

With the above lemmas holding for $\veps_n = o(a_n^{-1/2})$, lines for
proving 
Proposition 1 in~\cite{Li2017} can be followed  
exactly to finish the proof of Theorem \ref{thm:ACC_limit_small_bandwidth}. 


\section{Proof of Theorem 3}

\begin{lemma}\label{Alemma5} 
Assume Condition \ref{par_true}--\ref{cond:likelihood_moments}. If $\veps_{n}=o_{p}(a_{n}^{-3/5})$, then $a_{n}\veps_{n}(\beta_{\veps}-\beta_{0})=o(1)$.
\end{lemma}
\noindent {\it Proof of Lemma \ref{Alemma5}:} 
	This generalizes Lemma A4 in ~\cite{Li2017} by replacing $\pi(\theta_{0}+a_{n,\veps}^{-1}t)$
	therein with $\gamma_{n}(t)$. By Condition \ref{initial_upper} and the arguments in the proof
	of Lemma A4 in ~\cite{Li2017}, it can be shown that 
	\[
	\frac{q_{\mathbb{R}^{p}}\{(\theta-\theta_{0})^{k_{1}}(s-s_{\rm obs})^{k_{2}}\}}{q_{\mathbb{R}^{p}}(1)}=a_{n,\veps}^{-k_{1}}\veps_{n}^{-k_{2}}\left\{ \frac{\tilde{q}_{B_{\delta},tv}(t^{k_{1}}\nu^{k_{2}})}{\tilde{q}_{B_{\delta},tv}(1)}+O_{p}(\alpha_{n}^{-1})\right\} .
	\]
	Then by Lemma 2 $(iii)$, the right hand side of the above is equal
	to 
	\[
	a_{n,\veps}^{-k_{1}}\veps_{n}^{-k_{2}}\left\{ \frac{g_{B_{\delta},r}(t^{k_{1}}\nu^{k_{2}})}{g_{B_{\delta},r}(1)}+O_{p}(a_{n,\veps}^{-1})+O_{p}(a_{n}^{2}\veps_{n}^{4})+O_{p}(\alpha_{n}^{-1})\right\} .
	\]
	Since $\beta_{\veps}=\text{Cov}_{\veps}(\theta,S_{n})\text{Var}_{\veps}(S_{n})^{-1}$,
	\begin{align*}
	a_{n}\veps_{n}(\beta_{\veps}-\beta_{0})= & k_{n}\left[\frac{g_{B_{\delta},r}(t\nu)}{g_{B_{\delta},r}(1)}-\frac{g_{B_{\delta},r}(t)g_{B_{\delta},r}(\nu)}{g_{B_{\delta},r}(1)^{2}}+o_{p}(k_{n}^{-1})\right]\times\\
	& \qquad\left[\frac{g_{B_{\delta},r}(\nu\nu^{T})}{g_{B_{\delta},r}(1)}-\frac{g_{B_{\delta},r}(\nu)g_{B_{\delta},r}(\nu)^{T}}{g_{B_{\delta},r}(1)^{2}}+o_{p}(k_{n}^{-1})\right]^{-1}-a_{n}\veps_{n}\beta_{0},
	\end{align*}
	where the equations that $a_{n,\veps}^{-1}k_{n}=o(1)$, $a_{n}^{2}\veps_{n}^{4}k_{n}=o(p)$,
	and $\alpha_{n}^{-1}k_{n}=o(a_{n}^{-2/5}k_{n})=o(1)$ are used. By
	algebra, the right hand side of the equation above can be rewritten
	as 
	\begin{eqnarray*}
		&  & \left\{ \frac{g_{B_{\delta},r}\{(k_{n}t-a_{n}\veps_{n}\beta_{0}\nu)\nu\}}{g_{B_{\delta},r}(1)}-\frac{g_{B_{\delta},r}(k_{n}t-a_{n}\veps_{n}\beta_{0}\nu)g_{B_{\delta},r}(\nu)}{g_{B_{\delta},r}(1)^{2}}+o_{p}(1)\right\} \times\\
		&  & \qquad\left\{ E_{G,r}(\nu\nu^{T})-E_{G,r}(\nu)E_{G,r}(\nu)^{T}+o_{p}(k_{n}^{-1})\right\} ^{-1}.
	\end{eqnarray*}
	By plugging \eqref{eq6} and \eqref{eq7} in the above, $a_{n}\veps_{n}(\beta_{\veps}-\beta_{0})$
	is equal to 
	\begin{eqnarray*}
		&  & \left\{ E_{G,r}(\nu)\beta_{0}A(\theta_{0})^{1/2}W_{\rm obs}-E_{G,r}(\nu)\beta_{0}A(\theta_{0})^{1/2}W_{\rm obs}+o_{p}(1)\right\} \times \\
		&  & \hspace{1cm}
		\{\text{Var}_{G,r}(\nu)+o_{p}(k_{n}^{-1})\}^{-1}\\
		&=& o_{P}(1)\{\text{Var}_{G,r}(\nu)+o_{p}(k_{n}^{-1})\}^{-1}.
	\end{eqnarray*}
	Since
	\begin{align*}
	\text{Var}_{G,r}(\nu) & \geq\frac{\inf_{t\in t(B_{\delta'})}\gamma_{n}(t)}{g_{B_{\delta},r}(1)}\int_{\mathbb{R}^{d}}\int_{t(B_{\delta'})}\{\nu-E_{G,r}(\nu)\}^{2}g_{n}(t,\nu)\,dtd\nu,
	\end{align*}
	where $\delta'$ is defined in the proof of Lemma \ref{Alemma2}(ii),
	we have $\text{Var}_{G,r}(\nu)^{-1}=\Theta_{p}(1)$. Therefore $a_{n}\veps_{n}(\beta_{\veps}-\beta_{0})=o_{p}(1)$.
	\hfill{$\square$} 

\begin{lemma}\label{Alemma6} Results generalizing Lemma A5 in ~\cite{Li2017},
	$i.e.$ replacing $\Pi_{\veps}$ and $\tilde{\Pi}_{\veps}$ therein with
	$Q_{\veps}$ and $\tilde{Q}_{\veps}$, hold. \end{lemma} 
\noindent {\it Proof of Lemma \ref{Alemma6}:} 
	In ~\cite{Li2017}, the proof of Lemma A5 in ~\cite{Li2017} requires Lemma A4 and Lemma 2 in the supplements therein to hold. Since we have show that their generalized results hold for $\veps_{n}=o_{p}(a_{n}^{-1/2})$, (see Lemma \ref{Alemma5} and Lemma \ref{Alemma3.5} above),
    the proof of this lemma for $\veps_{n}=o_{p}(a_{n}^{-3/5})$ follows the same arguments in~\cite{Li2017}, replacing $\pi(\theta)$ with a $r_n(\theta)$ that satisfies conditions \ref{par_true} -- \ref{initial_gradient}.

 	\hfill{$\square$} 


	

With all above lemmas, the proof of Theorem \ref{thm:ACC_limit_large_bandwidth} holds by following the same arguments as those in the proof of Theorem 1 in ~\cite{Li2017}.

\bibliographystyle{nessart-number}
\bibliography{ACC_paper_2021}